\documentclass[3p]{elsarticle}

\usepackage{caption}
\usepackage{comment}
\usepackage{multirow}
\usepackage{tabularx}
\usepackage{colortbl}
\usepackage[ruled, lined, linesnumbered, commentsnumbered, longend]{algorithm2e}
\usepackage[most]{tcolorbox}
\usepackage{caption}
\usepackage{amssymb}
\usepackage{array}
\usepackage{diagbox}
\usepackage{hyperref}
\usepackage{supertabular}
\usepackage{booktabs}
\usepackage{float}
\usepackage{enumitem}
\usepackage{graphicx}
\usepackage{comment}
\usepackage{multirow}
\usepackage{tabularx}
\usepackage{colortbl}
\usepackage{longtable}
\usepackage{boldline}
\usepackage{float}
\usepackage{subcaption}
\usepackage{arydshln}
\journal{}

\begin{document}

\begin{frontmatter}

\title{Co-Tuning of Cloud Infrastructure and Distributed Data Processing Platforms}


\author{Isuru Dharmadasa}
\author{Faheem Ullah\corref{mycorrespondingauthor}}
\cortext[mycorrespondingauthor]{Corresponding author}

\address{The University of Adelaide, Australia}
\address{\{isuru.mahaganiarachchige, faheem.ullah\}@adelaide.edu.au}

\begin{abstract}
Distributed Data Processing Platforms (e.g., Hadoop, Spark, and Flink) are widely used to store and process data in a cloud environment. These platforms distribute the storage and processing of data among the computing nodes of a cloud. The efficient use of these platforms requires users to (i) configure the cloud i.e.,  determine the number and type of computing nodes, and (ii) tune the configuration parameters (e.g., data replication factor) of the platform. However, both these tasks require in-depth knowledge of the cloud infrastructure and distributed data processing platforms. Therefore, in this paper, we first study the relationship between the configuration of the cloud and the configuration of distributed data processing platforms to determine how cloud configuration impacts platform configuration. After understanding the impacts, we propose a co-tuning approach for recommending optimal co-configuration of cloud and distributed data processing platforms. The proposed approach utilizes machine learning and optimization techniques to maximize the performance of the distributed data processing system deployed on the cloud. We evaluated our approach for Hadoop, Spark, and Flink in a cluster deployed on the OpenStack cloud. We used three benchmarking workloads (WordCount, Sort, and K-means) in our evaluation. Our results reveal that, in comparison to default settings, our co-tuning approach reduces execution time by 17.5\% and \$ cost by 14.9\% solely via configuration tuning. 
\end{abstract}


\end{frontmatter}

% \fi
%\IEEEpeerreviewmaketitle

\section{Introduction}

\begin{figure}[h]
\captionsetup{justification=centering}
\centering
  \begin{subfigure}{0.7\textwidth}
  \centering
  \includegraphics[width=\linewidth]{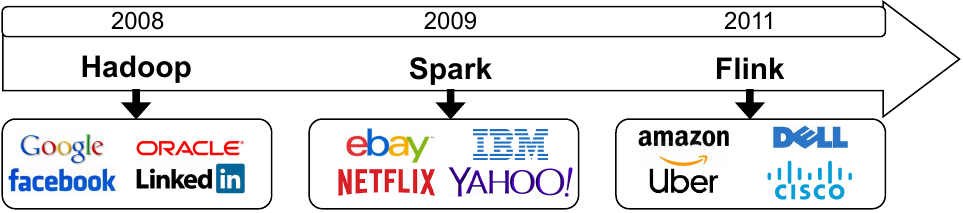}
\end{subfigure}
\caption{Initial release year and tech companies using Hadoop, Spark, and Flink.} 
\label{tech}
\vspace{-1.5 em}
\end{figure}

The increasing volume, velocity, and variety of data has led to the widespread adoption of cloud computing \cite{rimal2009taxonomy}. This is because the storage and processing of such a large volume of data is beyond the computational capacity of a single computer. Therefore, the cloud's high computational capacity is leveraged to deal with the large data size. Furthermore, the cloud capabilities are used in civil \cite{bello2021cloud, bhatt2016cloud} and tactical environments \cite{ahmad2023review, baek2021c4i}. In order to leverage the computing nodes of a cloud in an integrated way, distributed data processing platforms (hereafter referred as \textit{data platforms}) are used to store and process data on these nodes. These platforms facilitate users by hiding the internal complexities of how data is distributed for processing and results being merged after the completion of processing \cite{inoubli2018experimental}. The most commonly used data platforms are Hadoop, Spark, and Flink. Hadoop is one of the earliest platforms that follows the programming model of MapReduce \cite{dittrich2012efficient}. Hadoop is currently used by several tech industries such as Google, Facebook, and Oracle. Spark was introduced in 2009 to overcome the shortcomings of Hadoop such as in-memory processing. Spark is also quite famous among tech giants such as Amazon and NetFlix \cite{zaharia2016apache}. Flink is the most recent entry into the market that supports automated memory management, which was missing in Spark. Flink is also making its mark in several tech companies like Alibaba and Uber \cite{carbone2015apache}. Figure \ref{tech} shows the year each platform was introduced and some example tech companies currently using the platform.

%The cloud services are primarily available in two models - private and public. Private cloud refers to the internal data centers of an organization exclusively used and maintained by the respective organization or a third party on behalf of the organization \cite{rimal2009taxonomy}. Public cloud is used on a pay-as-you-go basis that is hosted and maintained by cloud providers such as Google, Microsoft, and Amazon \cite{bokhari2018survey}.

Both cloud and data platforms offer several configuration parameters that should be tuned to use them to their maximum potential \cite{herodotou2020survey}. For example, a cloud user needs to determine (i) what should be the number of computing nodes in the cluster (ii) what should be the type (e.g., small/large) of nodes, and (iii) whether or not to have all nodes of the same type. Similarly, the data platforms also come with several tunable parameters. As an example, a user needs to configure (i) the replication factor for each data block (ii) the number of executor cores allocated to a job, and (iii) the compression mode of the data during processing. The configurations of cloud infrastructure and data platforms directly impact the quality-of-service delivered by a distributed system deployed on a cloud. For example, the execution time and cost incurred by a distributed system with a sub-optimal configuration can be 10 times more than those when using optimal configuration \cite{hsu2018scout}. Therefore, it is important to tune both cloud and data platforms for optimal quality-of-service. However, optimal tuning of cloud and data platforms is a challenging task for multiple reasons. First, both cloud and data platforms have many parameters to control various aspects such as data replication factor. Consequently, it is too expensive to evaluate all combinations to determine the optimal configuration. Second, the optimal tuning of cloud and data platforms varies from workload to workload - a configuration optimal for one workload is not necessarily optimal for another. Third, the goal of tuning (e.g., reduce execution time, increase CPU utilization, improve scalability, etc.) may also vary depending upon user requirements, which complicates the process of tuning. In order to address these challenges, automated tuning techniques are required to take off the tuning burden from users' shoulders.

Several studies (e.g., \cite{marcu2016spark, zhu2017bestconfig, yigitbasi2013towards, wu2013self, wang2012predator, ullah2022scalability}) have investigated the impact of tuning configuration parameters and have proposed techniques for automated tuning. Marco et al. \cite{marcu2016spark} studied the impact of four Spark and four Flink parameters on the performance of the respective platforms. Similarly, in \cite{yigitbasi2013towards, wu2013self,wang2012predator}, the authors have explored the impact of tuning Hadoop parameters. On the other hand, several studies (e.g., \cite{herodotou2011no, lama2012aroma}) also investigated the impact of cloud configuration on the performance of distributed systems. For example, Herodotou et al. \cite{herodotou2011no} compared the performance of two cloud clusters - one consisting of homogeneous nodes and another consisting of heterogeneous nodes. Similarly, in \cite{lama2012aroma}, the authors explored whether a cluster consisting of several but smaller size nodes yields better performance or a cluster consisting of few but large size nodes yields better performance. In addition to exploratory studies, several studies (e.g., \cite{yigitbasi2013towards, perez2018pets}) also proposed solutions for automatically tuning data platforms. For example, Yigitbasi et al. \cite{yigitbasi2013towards} proposed a machine learning-based approach for tuning Hadoop. In \cite{perez2018pets}, the authors proposed a fuzzy-based approach for tuning Spark. Similarly, previous studies have proposed solutions for automatic tuning of cloud infrastructure. As an example, Zhu et al. \cite{zhu2017bestconfig} and Hsu et al. \cite{hsu2018scout} respectively, leveraged recursive bound-and-search and Bayesian optimization for automated tuning of cloud infrastructure.   

Previous studies have separately investigated either the impact of data platform parameters or cloud parameters. However, to the best of our knowledge, none of the studies have explored how tuning cloud parameters impact the tuning of data platform parameters. In other words, whether or not there is a relationship between the way we configure cloud and the way we configure data platform. For example, if we have 4 nodes in a cluster, whether we should tune executor cores of a data platform to 2 or 4 as compared to if we have 8 nodes. Furthermore, previous studies (e.g., \cite{zhu2017bestconfig, yigitbasi2013towards, wu2013self, wang2012predator, ullah2022adaptive}) have proposed tuning solutions either for tuning data platform configuration or cloud configuration. This way, either the potential of the data platform is fully exploited or the potential of the cloud. The previous studies have not proposed tuning solution that can co-tune both platform and cloud. In other words, there is a lack of a tuning solution that can tune the cloud and data platform at the same time to exploit the maximum potential of both the cloud and the data platform.

In this paper, we fill these two gaps by first investigating the impact of cloud configuration on data platform configuration. We study how the configuration of the cloud impacts the configuration of the platform. After understanding the impact, we propose an automated tuning approach that uses machine learning and a recursive random search algorithm for recommending near-optimal configuration for cloud and data platforms. We leverage the machine learning approach to capture complex system dynamics that other techniques (e.g., rule-based and simulation-based) are unable to capture \cite{herodotou2020survey}. Also, the use of machine learning enables us to utilize our learning obtained from real experiments reported in Section \ref{correlation}. Our tuning approach leverages the data generated as part of the exploratory study to understand the impact of cloud configuration on data platform configuration. We study the relationship between the configuration of data platform and cloud for the three most widely used platforms - Hadoop, Spark, and Flink. We have used three commonly used benchmarking workloads (Sort, Word Count, and K-means) in this study. Furthermore, we have used OpenStack as our cloud infrastructure where we deployed different types of Virtual Machines (VMs).\\

\textbf{\underline{Contributions}}: In this paper, we make the following contributions.
\begin{itemize}
    \item We investigate the relationship between the configuration of the cloud and the configuration of the data platform. We determine whether the way we configure the cloud impacts the way we configure the platform. 
    \item We propose a machine learning-based approach for co-tuning of cloud and data platform to exploit the full potential of both cloud and platform for maximizing the performance of the distributed data processing system. 
    \item We evaluate our tuning approach via rigorous experimentation using three bench-marking workloads and a large-scale cluster deployed on OpenStack. Our results reveal the effectiveness of our approach in terms of reducing execution time and cost. 
\end{itemize}

\textbf{\underline{Paper Structure}}: The rest of this paper is organized as follows. Section 2 provides an overview of the data platforms and cloud configuration. Section 2 also delineates the related work. Section 3 presents the experimental setup. Section 4 reports the findings about the relationship between cloud configuration and data platform configuration. Section 5 presents our co-tuning approach and its evaluation. Finally, Section 6 concludes the paper.

\section{Background and Related Work}
In this section, we introduce data platforms and their configurations, cloud configuration, and present related work to position the novelty of our work.
\subsection{Background}\label{background}

We provide an overview of the studied data platforms and cloud configurations in this section.
\subsubsection{Data platforms}\label{data_platforms}
We selected three data platforms for this study. These platforms are Hadoop, Spark, and Flink. This selection was based on the widespread popularity and adoption of these platforms in industry and academia \cite{oussous2018big, inoubli2018experimental}. The prominent features of Hadoop, Spark, and Flink are illustrated and compared in Table \ref{feature_comparison}. In the following, we discuss and contrast these features of Hadoop, Spark, and Flink.

\textbf{Hadoop:} Hadoop is one of the earliest data platforms introduced in 2008 \cite{polato2014comprehensive}. It is an open-source project managed by Apache Foundation. Hadoop primarily consists of two parts - MapReduce and Hadoop Distributed File System (HDFS). MapReduce is the programming model that consists of map and reduce functions, which divides data into key-value pairs. HDFS is the file system, which Hadoop uses for reading and writing data. Unlike Spark and Flink, Hadoop is a disk-based platform that shuffles data between disk and memory during data processing. A study conducted at the University of Sydney concluded that Hadoop is not as user-friendly as Spark and Flink \cite{akil2017usability}. Furthermore, Hadoop offers little flexibility in terms of programming languages as the software to run on Hadoop is mainly written in Java.

\textbf{Spark:} Spark is also an Apache Foundation project that was initially introduced in 2009 at UC Berkeley \cite{zaharia2010spark}. The aim behind the development of Spark was to overcome the shortcomings of Hadoop such as slower processing speed. Over the years, Spark has developed into a matured data processing platform incorporated by several tech giants such as Yahoo, eBay, and IBM. Spark relies on Resilient Distributed Datasets (RDD), which are essentially data objects on which two types of operations (transformation and actions) are performed during data processing. Unlike Hadoop, Spark brings all data into memory for processing. Once the data processing is completed, the results are written back to the disk. Spark offers flexibility in terms of coding languages. Also, Spark brings all configuration parameters into one file, hence, making parameter tuning comparatively easier.

\begin{table}[]
\centering
\caption{Feature-wise comparison of Hadoop, Spark, and Flink}
\scriptsize
%\resizebox{\textwidth}{!}{%
\begin{tabular}{l|l|l|l}
\hlineB{3}
\textbf{Feature} & \textbf{Hadoop} & \textbf{Spark} & \textbf{Flink} \\ \hlineB{3}
\textbf{Data processing  mode} & Batch & Batch and stream & Batch and stream \\
\textbf{Data source} & HDFS & HDFS, DBMS, Kafka & HDFS, Kafka, Message queue \\
\textbf{Data format} & Key-value & Key-value, RDD & Key-value \\
\textbf{Data processing model} & Disk-based & RAM-based & RAM-based \\
\textbf{Usability} & Difficult to use & Easy to use & Easy to use \\
\textbf{Memory management} & Manual & Manual & Automatic \\
\textbf{Supporting languages} & Java & Java, Scala, Python, R & Java, Scala, Python, R \\
\textbf{Job optimization} & Manually optimized & Manually optimized & In-built optimizer \\
\textbf{Data caching} & No-caching & Cache data in memory & Cache data in memory \\
\textbf{Cost} & Lower cost & Higher cost & Higher cost \\
\textbf{Cluster manager} & YARN & YARN, Mesos, standalone & YARN, standalone, Zookeepr \\
\textbf{Programming model} & Map and Reduce & Transformation and Action & Transformation \\
\textbf{ML library} & Mahout & SparkML and MLlib & FlinkML \\ \hlineB{3}
\end{tabular}%
%}
\label{feature_comparison}
\end{table}

\textbf{Flink:} Flink was introduced in 2011 to overcome the shortcomings of Spark, especially the ones related to optimizations \cite{carbone2015apache}. Similar to Spark, Flink is designed both for batch and stream processing. However, it has a significantly different architecture that offers its own benefits. For instance, the allocation of memory for running a data processing job is automatically configured in Flink, which leads to yield better performance. In addition, Flink has its built-in optimizer that further boosts the performance. However, since Flink is not as mature as Hadoop and Spark, its ML library (FlinkML\footnote{https://bit.ly/3wcZimU}) offers very few choices for machine learning tasks. The programming model (MapReduce) followed by Flink is similar to that of Hadoop.

\textbf{Configuring Hadoop, Spark, and Flink:} Like any large-scale software system, these data platforms also need to be tuned in order to use them to the maximum of their ability. For example, based on their performance studies, both Babu et al. \cite{babu2010towards} and Jiang et al. \cite{jiang2010performance} concluded that the tuning of Hadoop parameters can significantly improve Hadoop's execution time. Tuning these platforms not only help reduce the execution time but also directly impacts several other quality attributes such as scalability, reliability, resource utilization and energy consumption. For example, replication factor is a parameter that users need to tune because the number of nodes on which a data block is copied directly impacts the reliability of the system. Consider a scenario where the data is available only on one node and that node goes down. All operations depending on that node's data will fail. Similarly, determining whether or not to compress the output of map function before forwarding it to the reduce function directly impacts the execution time. Hence, tuning these platforms to maximize the gain is an important task. However, each of these platforms has hundreds of configuration parameters/tuning knobs, which makes tuning a tedious and time-consuming task. 

Not all parameters of data platforms significantly impact the performance of a distributed system \cite{herodotou2020survey}. Therefore, it is important to identify the parameters that significantly influence performance. These parameters are often related to data compression, job scheduling, and resource allocation. Previous studies have selected parameters for tuning based on expert's opinion \cite{wang2016novel}, industry guidelines\footnote{https://sparkhub.databricks.com/}, and studies \cite{vaquero2018auto}, \cite{ yang2012statistics} that have established a strong correlation between the parameters and system's performance. In light of these sources, we also select parameters of Hadoop, Spark, and Flink for consideration in this study. These parameters along with their potential values are presented in Table \ref{hadoop_parameters_table}, Table \ref{spark_parameters_table}, Table \ref{flink_parameters_table}, respectively. These are the parameters found most important by previous studies as well as industrial practices \cite{herodotou2020survey}, \cite{costa2021survey}. The selection of potential values for Boolean and categorical parameters is straightforward. This is because Boolean parameters, such as \textit{output.fileoutputformat.compress} has only two possible values i.e., TRUE and FALSE. Similarly, the potential values for categorical parameters like \textit{output.fileoutputformat.codec} have been already defined. We selected the potential values for numerical parameters, such as \textit{tasktracker.map.tasks.maximum}, based on the previous studies \cite{herodotou2020survey}, \cite{costa2021survey}.  

Apart from the challenge posed by the large number of parameters to tune, the configuration tuning also varies with respect to various factors such as workload and the platform itself. For example, Petridis et al. \cite{petridis2016spark} found that, in comparison to default settings, tuning \textit{spark.rdd.compress} to FALSE in Spark reduces execution time by 5\% for machine learning tasks but increases the execution time by 5\% for sorting. Similarly, optimal tuning from one platform does not directly transfer to another platform. For instance, the study by Marcu et al. \cite{marcu2016spark} revealed that doubling the number of executor cores improves the execution time of Flink but reduces the execution time of Spark. Hence, configuring tuning of data platforms is crucial but a tedious and complex task. 

\begin{table}[t]
\caption{Studied Hadoop parameters, their description, and search space with default value (A) and potential values B, C, and D.}
\resizebox{\textwidth}{!}{%
\begin{tabular}{l|l|l|l}
\hlineB{3}
\textbf{ID} & \textbf{Hadoop Parameter} & \textbf{Description} & \textbf{Search Space \{A, B, C, D\}} \\ \hlineB{3}
H1 & output.fileoutputformat.compress & Should final output be compressed & \{FALSE, TRUE\} \\ 
H2 & output.fileoutputformat.compress.type & Type of compression for the final output & \{RECORD, BLOCK\} \\ 
H3 & output.fileoutputformat.compress.codec & Name of codec in case compression for final output is enabled & \{Default, Gzip, Bzip2, Lz4\} \\
H4 & output.compress & Whether or not to compress the output of map phase & \{TRUE, FALSE\} \\
H5 & map.output.compress.codec & Name of codec in case compression for map output is enabled & \{Default, Gzip, Bzip2, Lz4\} \\ 
H6 & tasktracker.map.tasks.maximum & Maximum number of map tasks executed in parallel & \{2, 1, 4\} \\ 
H7 & tasktracker.reduce.tasks.maximum & Maximum number of reduce tasks executed in parallel & \{2, 1, 4\} \\ 
H8 & child.java.opts & Determines memory size allocation for a task & \{-Xmx200m, -Xmx1639m\} \\ 
H9 & map.speculative & Whether or not to execute some map instances in parallel & \{TRUE, FALSE\} \\ 
H10 & reduce.speculative & Whether or not to execute some reduce instances in parallel & \{TRUE, FALSE\} \\
H11 & task.io.sort.mb & Amount of buffer memory to use for sorting & \{100, 1150\} \\ 
H12 & task.io.sort.factor & How many streams should merge together at once during sorting & \{10, 100\} \\ 
H13 & map.sort.spill.percent & Determines the limit for the serialization buffer & \{0.8, 0.3\} \\ \hlineB{3}
\end{tabular}%
}
\label{hadoop_parameters_table}
\end{table}

\begin{table}[t]
\caption{Studied Spark parameters, their description, and search space with default value (A) and potential values B, and C.}
\resizebox{\textwidth}{!}{%
\begin{tabular}{l|l|l|l}
\hlineB{3}
\textbf{ID} & \textbf{Spark Parameter}     & \textbf{Description}                                                                            & \textbf{Search Space \{A, B, C\}}      \\ \hlineB{3}
S1          & io.compression.coded         & Type of codec for compression of internal data                                                  & \{Iz4, Izf, Snappy\}                 \\ 
S2          & serializer                   & How to serialize objects                                                                        & \{Java Serializer, Kryo Serializer\} \\ 
S3          & io.compression.Iz4.blockSize & Size of block in lz4 codec                                                                      & \{32k, 16k, 64k\}                    \\ 
S4          & shuffle.spill.compress       & Whether or not to compress outputs of map                                                       & \{TRUE, FALSE\}                      \\ 
S5          & reducer.maxSizeInFlight      & Amount of maximum map output being fetched                                                      & \{48m, 24m, 72m\}                    \\ 
S6          & shuffle.file.buffer          & Amount of in-memory buffer size in shuffles                                                     & \{32k, 16k, 48k\}                    \\ 
S7          & shuffle.compress             & Whether or not to compress outputs of map                                                       & \{TRUE, FALSE\}                      \\ 
S8          & broadcast.blockSize          & Size of block piece                                                                             & \{4m, 2m, 8m\}                       \\ 
S9          & locality.wait                & \begin{tabular}[c]{@{}l@{}}Amount of time being waited for launching local tasks\end{tabular} & \{3s, 1s, 5s\}                       \\ 
S10         & memory.fraction              & Fraction of memory for execution and storage                                                    & \{0.6, 0.4, 0.8\}                    \\
S11         & memory.storageFraction       & Amount of memory for storage without eviction                                                   & \{0.5, 0.3, 0.7\}                    \\ \hlineB{3}
\end{tabular}%
}
\label{spark_parameters_table}
\end{table}

\begin{table}[t]
\caption{Studied Flink parameters, their description, and search space with default value (A) and potential values B, C, D, and E.}
\resizebox{\textwidth}{!}{%
\begin{tabular}{l|l|l|l}
\hlineB{3}
\textbf{ID} & \textbf{Flink Parameter}                     & \textbf{Description}                            & \textbf{Search Space \{A, B, C, D, E\}} \\ \hlineB{3}
F1          & memory.managed.fraction                      & Fraction of memory used as managed memory       & \{0.4, 0.2, 0.3, 0.5, 0.6\}               \\ 
F2          & memory.jvm-overhead.fraction                 & Fraction of memory reserved for JVM overhead    & \{0.1, 0.05, 0.2\}              \\ 
F3          & memory.network.fraction                      & Fraction memory used as network memory          & \{0.1, 0.05, 0.2\}              \\ 
F4          & network.blocking-shuffle.compression.enabled & Whether or not to compress shuffle data         & \{FALSE, TRUE\}                 \\ 
F5          & network.memory.buffers-per-channel           & Number of network buffers for outgoing/incoming & \{2,4,6\}                       \\ 
F6          & network.netty.server.numThreads              & Number of Netty server threads                  & \{-1, 1, 2\}                    \\ 
F7          & network.netty.clinet.num-threads             & Number of Netty client threads                  & \{-1, 1, 2\}                    \\ 
F8          & execution.checkpointing.snapshot-compression & Whether or not to compress snapshot data        & \{FALSE, TRUE\}                 \\ \hlineB{3}
\end{tabular}%
}
\label{flink_parameters_table}
\end{table}

\subsubsection{Cloud configurations}
Whilst cloud computing offers several advantages for distributed data analytics, using cloud resources in a timely and cost-efficient manner is not a straightforward task, especially for non-expert users. Consequently, users suffer not only from a lack of performance guarantee but also from avoidable costs. In order to guarantee performance and avoid extra costs, cloud resources need to be properly configured. Such configuration involves sizing the cloud cluster that requires users to determine the optimal number of nodes and selecting the type of nodes from several available choices. For example, some node types and their features offered by well-known cloud providers (i.e., Amazon, Google, and Microsoft) are shown in Table \ref{amazon_azure_google} \cite{amazon_azure_google_ref}. For instance, looking at Table \ref{amazon_azure_google}, a user needs to decide whether to choose one Amazon node of \textit{r3.large} type or two nodes each of \textit{m3.large} type. Similar to the tuning of data platforms, the configuration of the cloud directly impacts several quality attributes such as execution time, cost, and scalability. However, at the same time, optimal tuning of the cloud is also a challenging task due to multiple reasons such as the large pool of potential configurations and variation in optimal cloud configuration as per the change in the workload. For example, Alipourfard et al. \cite{alipourfard2017cherrypick} demonstrated that the \$ cost for regression decreases as the RAM/core ratio in a cloud increases. In contrast, the \$ cost for sorting increases as the RAM/core ratio increases. Therefore, similar to data platforms, configuring the cloud is an important but challenging task.

\begin{table}[t]
\caption{Representative instance/node types offered by Amazon, Azure, and Google along with their RAM, disk storage, and cost/hour.}
\scriptsize
\begin{tabular}{l|l|l|l|l|l|l|l|l|l|l|l}
\hlineB{3}
\multicolumn{4}{c|}{\textbf{Amazon}}     & \multicolumn{4}{c|}{\textbf{Azure}}   & \multicolumn{4}{c}{\textbf{Google}}         \\ \hlineB{3}
Type     & RAM   & Disk & Cost & Type   & RAM & Disk & Cost & Type          & RAM & Disk & Cost \\ \hline
m3.large & 8 GB     & 32 GB   & \$0.133   & D2 v2  & 7 GB   & 100 GB  & \$0.114   & n1-standard-2 & 7.5 GB & 375 GB  & \$0.212   \\ 
r3.large & 15 GB    & 32 GB   & \$0.166   & D11 v2 & 14 GB  & 100 GB  & \$0.149   & n1-highmem-2  & 13 GB  & 375 GB  & \$0.018   \\ 
c3.large & 3.75 GB  & 32 GB   & \$0.105   & F2     & 4 GB   & 32 GB   & \$0.099   & n1-highcpu-2  & 1.8 GB & 375 GB  & \$0.104   \\ 
r4.large & 15.2 GB & 0       & \$0.133   & D11 v2 & 14 GB  & 100 GB  & \$0.149   & n1-highment-2 & 13 GB  & 0       & \$0.010   \\ 
c4.large & 3.75 GB  & 0       & \$0.105   & F2     & 4 GB   & 32 GB   & \$0.099   & n1-highcpu-2  & 1.8 GB & 0       & \$0.076   \\ \hlineB{3}
\end{tabular}
\label{amazon_azure_google}
\end{table}

\subsection{Related Work}
We divide the related works into two streams - first reporting studies on tuning of data platforms and the second stream reports studies on tuning of cloud. In order to position the novelty of our work, the key features of the related works are depicted in Table \ref{related_work}.

\begin{table}[]
\caption{Comparison of previous studies with our work. Tuning relationship refers to the correlation between the tuning of data platform and tuning of cloud. Co-tuning refers to tuning both data platform and cloud at the same time}
\resizebox{\textwidth}{!}{%
\begin{tabular}{l|l|l|l|l|l|l}
\hlineB{3}
\textbf{Study} & \textbf{Studied Platform} & \textbf{Platform Tuning} & \textbf{Cloud Used} & \textbf{Cloud Tuning} & \textbf{Tuning Relationship} & \textbf{Co-Tuning} \\ \hlineB{3}
Yigitbasi et al. \cite{yigitbasi2013towards} & Hadoop & Yes & Private cloud & No & No & No \\
Wang et al. \cite{wang2016novel} & Spark & Yes & Private cloud & No & No & No \\
Wu and Gokhale. \cite{wu2013self} & Hadoop & Yes & Amazon EC2 & No & No & No \\
Wang et al. \cite{wang2012predator} & Hadoop & Yes & Private cloud & No & No & No \\
Herodotou et al. \cite{herodotou2011starfish} & Hadoop & Yes & Amazon EC2 & No & No & No \\ 
Perez et al. \cite{perez2018pets} & Spark & Yes & Private cloud & No & No & No \\
Zhu et al. \cite{zhu2017bestconfig} & Spark & Yes & Huawei & No & No & No \\ 
Yu et al. \cite{yu2018datasize} & Spark & Yes & Private cloud & No & No & No \\ 
Gounaris et al. \cite{gounaris2018methodology} & Spark & Yes & Private cloud & No & No & No \\ 
Herodotou et al. \cite{herodotou2011no} & Hadoop & No & Amazon EC2 & Yes & No & No \\ 
Chen et al. \cite{chen2013cresp} & Hadoop & No & Private & Yes & No & No \\
Fu et al. \cite{fu2015drs} & Storm & No & Private cloud & Yes & No & No \\ 
Lama et al. \cite{lama2012aroma} & Hadoop & No & Private cloud & Yes & No & No \\
Lee et al. \cite{lee2011heterogeneity} & Hadoop & No & Amazon EC2 & Yes & No & No \\
\textbf{This work} & \textbf{Hadoop, Spark, Flink} & \textbf{Yes} & \textbf{Private cloud} & \textbf{Yes} & \textbf{Yes} & \textbf{Yes} \\ \hlineB{3}
\end{tabular}%
}
\label{related_work}
\end{table}

Several studies have explored and/or proposed techniques for tuning data platforms. In \cite{yigitbasi2013towards, wang2016novel}, the authors proposed machine learning-based approaches for tuning Hadoop and Spark, respectively. For tuning Hadoop, Wu et al. \cite{wu2013self} combined K-means with simulated annealing to learn about configurations and search for optimal configurations for the unknown Hadoop jobs. Both Predator \cite{wang2012predator} and Starfish \cite{herodotou2011starfish} leverage cost model and profiling information to obtain global optimum for configuration tuning of Hadoop. However, unlike Starfish \cite{herodotou2011starfish}, Predator \cite{wang2012predator} combined practical Hadoop experience with the Hill Climbing algorithm to accurately and quickly converge towards optimal Hadoop configuration for a given job. Perez et al. \cite{perez2018pets} leveraged fuzzy logic for tuning 12 parameters of Spark. In \cite{zhu2017bestconfig}, the authors used a search-based optimization technique that exploits previously obtained performance models to tune various systems including Spark. Along the same lines, Zhibin et al. \cite{yu2018datasize} combined hierarchical modeling with genetic algorithms to search for optimal configuration for tuning Spark. Gounaris et al. \cite{gounaris2018methodology} developed a trial-and-error technique for tuning Spark, which tries various configuration settings in order to converge towards an optimal setting. All the above studies are solely focused on the tuning of data platforms without paying any attention to the configuration of the underlying cloud on which the platform is deployed. For further studies on tuning data platforms, interested readers are referred to \cite{herodotou2020survey}, \cite{costa2021survey}.

Similar to data platforms, several studies have investigated the configuration tuning of cloud infrastructure. Herodotou et al. \cite{herodotou2011no} proposed Elastisizer - a tuning system that receives the user's cloud requirements in the form of a query and recommends optimal cloud configuration in terms of the number and type of VMs. Elastisizer uses job profiling, estimation models, and simulations for recommending optimal cloud configuration. In \cite{chen2013cresp}, the authors studied the relationship among the data size, the complexity of the algorithm, and the available system resources to optimally allocate cloud resources for Hadoop.  In a similar study \cite{fu2015drs}, the authors established a relationship among cloud resources, user's requirements, and workload to optimally provision cloud resources for real-time data analytical jobs using Storm (used for stream data processing unlike batch processing as carried out in our work). However, the parameter tuning of the data platform and their relation with cloud resources are left out in both of these studies. Lama et al. \cite{lama2012aroma} proposed a technique for the automated provisioning of cloud resources for Hadoop jobs. The proposed technique leverages machine learning and optimization techniques for recommending cloud configuration such that the jobs are processed within predefined deadlines. Lee at el. \cite{lee2011heterogeneity} proposed the concept of resource sharing in heterogeneous cloud environments to optimize the usage of cloud resources. However, the proposed approach merely focuses on storage requirements without offering any performance guarantee. All the above studies are primarily focused on the configuration of cloud infrastructure without paying attention to the tuning of the applications (e.g., Hadoop) running on top of the cloud infrastructure. 

The focal point of the previous studies is either the tuning of data platforms or tuning of cloud infrastructure. Unlike the previous studies, this paper first determines the relationship between the tuning of data platform and the tuning of cloud. It then focuses on tuning both data platform and cloud at the same time to exploit the maximum potential of both the data platform and the cloud.

\section{Experimental Setup}\label{experimental_setup}
In this section, we describe the testbed, configurations of cloud and data platforms, and workloads used in this study. 

\subsection{Testbed} We used OpenStack\footnote{https://www.openstack.org/} private cloud platform, located at the University of Adelaide, to set up the cloud infrastructure for this study. Since the data platforms follow master-slave architecture, we chose \textit{m1.large} flavor (having 4 vCPUs, 80 GB disk, and 8 GB RAM) for implementing the master node of the cluster. For implementing slave/worker nodes, we used different node flavors presented in Table \ref{cloud_configurations_table} and discussed in Section \ref{cloud_platform_configurations}. We installed Ubuntu 16.04.7 LTS operating system on each node. Furthermore, we installed Java OpenJDK 1.8.0, Python2, MatPlotLib, and Maven on each node to facilitate the installation of Hadoop, Spark, and Flink. We used dstat\footnote{https://linux.die.net/man/1/dstat} tool for measuring CPU, RAM, and disk usage. 

\subsection{Configurations of Cloud and data platforms}\label{cloud_platform_configurations} With respect to the cloud configuration, our setup consists of eight types of worker nodes, which are divided into two broader categories i.e., medium denoted by ‘m.’ and large denoted by ‘l.’ as presented in Table \ref{cloud_configurations_table}. Within each category, there are four types - small, medium, large, and xlarge. This way \textit{m.small} is the node with the least resources and \textit{l.xlarge} is the node with the most resources. Based on these worker types, we designed 11 cloud configurations as presented in Table \ref{cloud_configurations_table}. These cloud configurations differ from each other in terms of the number and/or type of worker nodes in each configuration. For example, cloud configuration \textit{C0} consists of two nodes each of type \textit{l.small}. In contrast, cloud configuration \textit{C1} consists of two nodes - one of type \textit{m1.medium} and another of type \textit{l1.xlarge}. To have a fair comparison, we made sure that the total capacity of each cloud configuration is the same i.e., vCPUs (10), disk (200 GB), and RAM (20 GB). The parameters selected for each platform (Hadoop, Spark, and Flink) and their value options are presented in Table \ref{hadoop_parameters_table}, Table \ref{spark_parameters_table}, and Table \ref{flink_parameters_table} and previously discussed in Section \ref{data_platforms}. 

\subsection{Workloads} We used both batch and iterative workloads in this study. Batch workloads follow the principle of one-pass processing, where the input is processed only once. On the other hand, in iterative workload, the same input is processed multiple times in an iterative manner before producing the final output. In batch category, we selected Sort and Word Count while in iterative category, we selected K-means. \textit{Sort} is a workload that implements sorting algorithm, which can reasonably assess the I/O and communication performance of data platforms. \textit{Word Count} is a workload that counts the number of times a word appears in a text file. It can reasonably assess the aggregation capability of data platforms. \textit{K-means} implements the well-known algorithm for clustering data, which is a good choice for assessing the machine learning capabilities of data platforms. All three workloads are selected based on their widespread use in the related studies \cite{herodotou2020survey}, \cite{costa2021survey}. We used HiBench \cite{huang2010hibench} and BigDataBecnh \cite{wang2014bigdatabench} suites for the implementation of these workloads on Hadoop, Spark, and Flink.

\subsection{Experimental Execution} Our experimental execution requires changes with respect to various variables such as cloud configuration, platform configuration, the platform itself, and workloads. Therefore, during the experiment, we would change only one variable at a time with respect to default settings. For example, when Word Count workload was executed with Flink, we iteratively updated its \textit{memory.managed.fraction} parameter value from its default value 0.4 to 0.3 and 0.5. Subsequently, when we intended to investigate the impact of another parameter (e.g., \textit{memory.jvm-overhead.fraction)}, we made sure to set the previously investigated \textit{memory.managed.fraction} to its default value. Therefore, in this study, given any instance, only one variable value was altered compared to its default configuration setting. We modified \textit{mapred-site.xml}, \textit{spark-default.conf} and \textit{flink-conf.yaml} files to change configuration settings of Hadoop, Spark, and Flink, respectively.

\begin{table}[]
\caption{Cloud configurations used in this study}
\scriptsize
\resizebox{\textwidth}{!}{%
\begin{tabular}{l|ccc|ccccccccccc}
\hlineB{3}
\multirow{2}{*}{\textbf{Node Type}} & \multicolumn{3}{c|}{\textbf{Resources Specification}} & \multicolumn{11}{c}{\textbf{Cloud Configurations}} \\ \cline{2-15} 
 & \multicolumn{1}{c|}{vCPUs} & \multicolumn{1}{c|}{Disk (GB)} & RAM (GB) & \multicolumn{1}{c|}{C0} & \multicolumn{1}{c|}{C1} & \multicolumn{1}{c|}{C2} & \multicolumn{1}{c|}{C3} & \multicolumn{1}{c|}{C4} & \multicolumn{1}{c|}{C5} & \multicolumn{1}{c|}{C6} & \multicolumn{1}{c|}{C7} & \multicolumn{1}{c|}{C8} & \multicolumn{1}{c|}{C9} & C10 \\ \hline
m.small & \multicolumn{1}{c|}{1} & \multicolumn{1}{c|}{10} & 2 & \multicolumn{1}{c|}{} &  \multicolumn{1}{c|}{} & \multicolumn{1}{c|}{} & \multicolumn{1}{c|}{} & \multicolumn{1}{c|}{} & \multicolumn{1}{c|}{} & \multicolumn{1}{c|}{} & \multicolumn{1}{c|}{} & \multicolumn{1}{c|}{} & \multicolumn{1}{c|}{} & 2 \\ \hdashline
m.medium & \multicolumn{1}{c|}{2} & \multicolumn{1}{c|}{40} & 4 & \multicolumn{1}{c|}{} & \multicolumn{1}{c|}{1} & \multicolumn{1}{c|}{} & \multicolumn{1}{c|}{} & \multicolumn{1}{c|}{1} & \multicolumn{1}{c|}{} & \multicolumn{1}{c|}{1} & \multicolumn{1}{c|}{3} & \multicolumn{1}{c|}{2} & \multicolumn{1}{c|}{5} & 2 \\ \hdashline
m.large & \multicolumn{1}{c|}{3} & \multicolumn{1}{c|}{60} & 6 & \multicolumn{1}{c|}{} & \multicolumn{1}{c|}{} & \multicolumn{1}{c|}{1} & \multicolumn{1}{c|}{} & \multicolumn{1}{c|}{} & \multicolumn{1}{c|}{2} & \multicolumn{1}{c|}{1} & \multicolumn{1}{c|}{} & \multicolumn{1}{c|}{2} & \multicolumn{1}{c|}{} &  \\ \hdashline
m.xlarge & \multicolumn{1}{c|}{4} & \multicolumn{1}{c|}{80} & 8 & \multicolumn{1}{c|}{} & \multicolumn{1}{c|}{} & \multicolumn{1}{c|}{} & \multicolumn{1}{c|}{1} & \multicolumn{1}{c|}{2} & \multicolumn{1}{c|}{1} & \multicolumn{1}{c|}{} & \multicolumn{1}{c|}{1} & \multicolumn{1}{c|}{} & \multicolumn{1}{c|}{} & 1 \\ \hdashline
l.small & \multicolumn{1}{c|}{5} & \multicolumn{1}{c|}{100} & 10 & \multicolumn{1}{c|}{2} & \multicolumn{1}{c|}{} & \multicolumn{1}{c|}{} & \multicolumn{1}{c|}{} & \multicolumn{1}{c|}{} & \multicolumn{1}{c|}{} & \multicolumn{1}{c|}{1} & \multicolumn{1}{c|}{} & \multicolumn{1}{c|}{} & \multicolumn{1}{c|}{} &  \\ \hdashline
l.medium & \multicolumn{1}{c|}{6} & \multicolumn{1}{c|}{120} & 12 & \multicolumn{1}{c|}{} & \multicolumn{1}{c|}{} & \multicolumn{1}{c|}{} & \multicolumn{1}{c|}{1} & \multicolumn{1}{c|}{} & \multicolumn{1}{c|}{} & \multicolumn{1}{c|}{} & \multicolumn{1}{c|}{} & \multicolumn{1}{c|}{} & \multicolumn{1}{c|}{} &  \\ \hdashline
l.large & \multicolumn{1}{c|}{7} & \multicolumn{1}{c|}{140} & 14 & \multicolumn{1}{c|}{} & \multicolumn{1}{c|}{} & \multicolumn{1}{c|}{1} & \multicolumn{1}{c|}{} & \multicolumn{1}{c|}{} & \multicolumn{1}{c|}{} & \multicolumn{1}{c|}{} & \multicolumn{1}{c|}{} & \multicolumn{1}{c|}{} & \multicolumn{1}{c|}{} &  \\ \hdashline
l.xlarge & \multicolumn{1}{c|}{8} & \multicolumn{1}{c|}{160} & 16 & \multicolumn{1}{c|}{} & \multicolumn{1}{c|}{1} & \multicolumn{1}{c|}{} & \multicolumn{1}{c|}{} & \multicolumn{1}{c|}{} & \multicolumn{1}{c|}{} & \multicolumn{1}{c|}{} & \multicolumn{1}{c|}{} & \multicolumn{1}{c|}{} & \multicolumn{1}{c|}{} &  \\ \hlineB{3}
\end{tabular}%
}
\label{cloud_configurations_table}
\end{table}

\section{Impact of Cloud Configuration on Data Platform Configuration}\label{correlation}

In this section, we study the impact of cloud configuration on data platform configuration. In other words, we determine whether or not a relationship exists between cloud configuration and data platform configuration. 

\subsection{Correlation between Cloud Configuration on Hadoop Configuration}

\begin{figure*}[t]
\centering
\begin{subfigure}{.32\textwidth}
  \centering
  \includegraphics[trim=0 0 15 10, clip, scale = 0.34]{   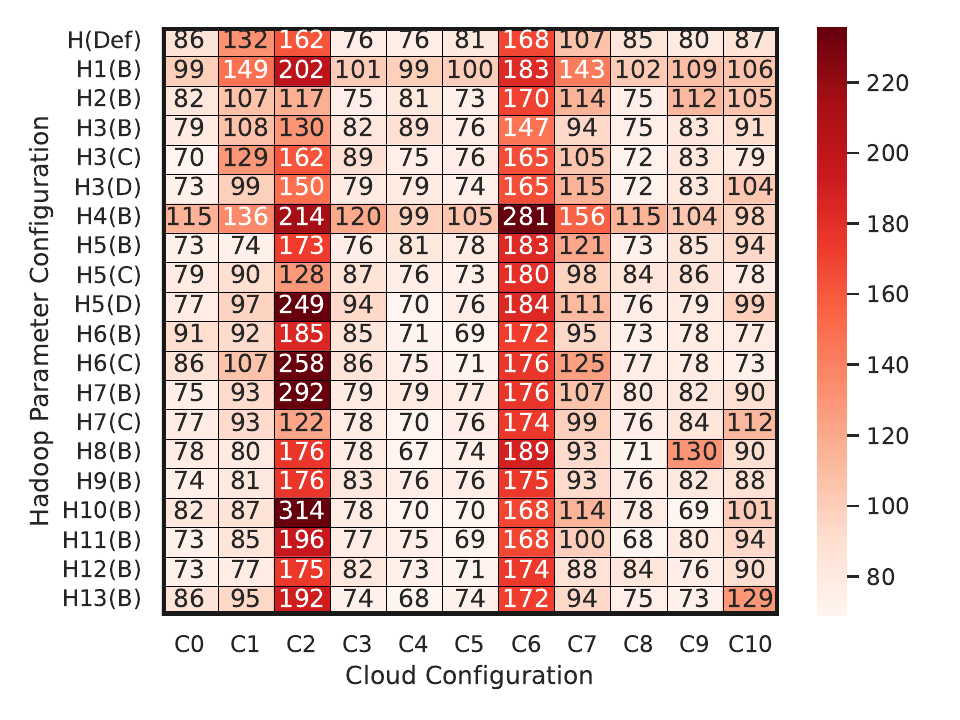}
  \caption{Sort}
  \label{h_sort_time}
\end{subfigure}%
\begin{subfigure}{.32\textwidth}
  \centering
  \includegraphics[trim=0 0 15 10, clip, scale = 0.34]{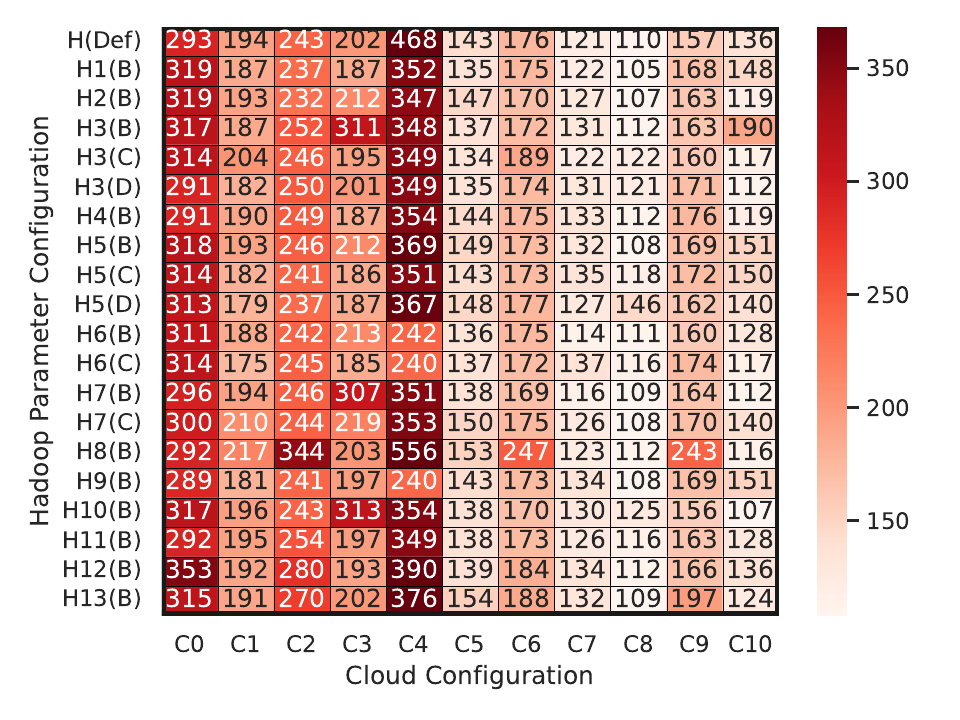}
  \caption{Word Count}
  \label{h_wordcount_time}
\end{subfigure}
\begin{subfigure}{.32\textwidth}
  \centering
  \includegraphics[trim=0 0 15 10, clip, scale = 0.34]{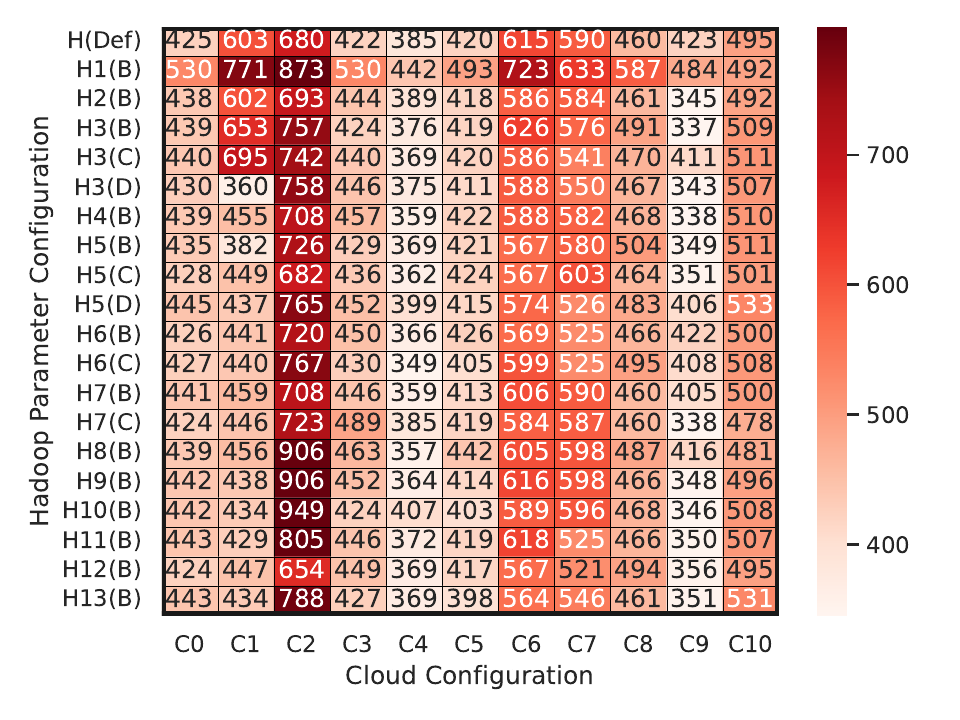}
  \caption{K-means}
  \label{h_kmeans_time}
\end{subfigure}
\caption{Execution time (in seconds) of \textbf{Hadoop} with respect to various cloud configurations and Hadoop parameter configurations for executing Sort, Word Count, and K-means. \textit{C0}, \textit{C1}, and so on (x-axis) specify the cloud configurations presented in Table \ref{cloud_configurations_table}. \textit{H(n)} on y-axis denotes Hadoop parameter configurations reported in Table \ref{hadoop_parameters_table}.}
\label{hadoop_execution_time}
\end{figure*}

\begin{figure*}[t]
\centering
\begin{subfigure}{.32\textwidth}
  \centering
  \includegraphics[trim=10 10 10 10, clip, scale = 0.33]{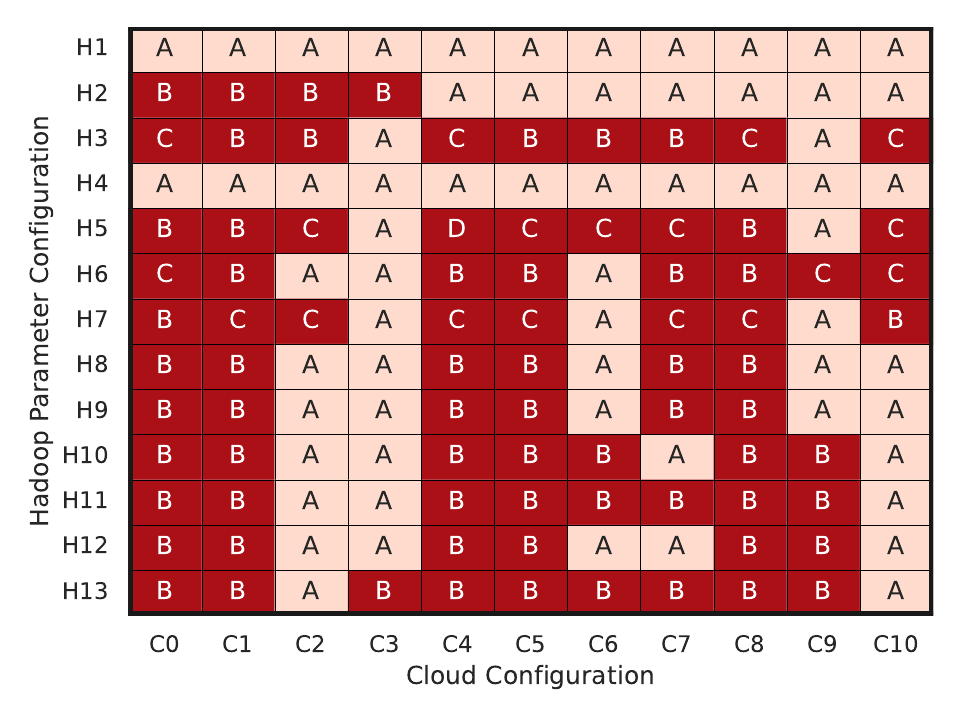}
  \caption{Sort}
  \label{hh_sort_time}
\end{subfigure}%
\begin{subfigure}{.32\textwidth}
  \centering
  \includegraphics[trim=10 10 10 10, clip, scale = 0.33]{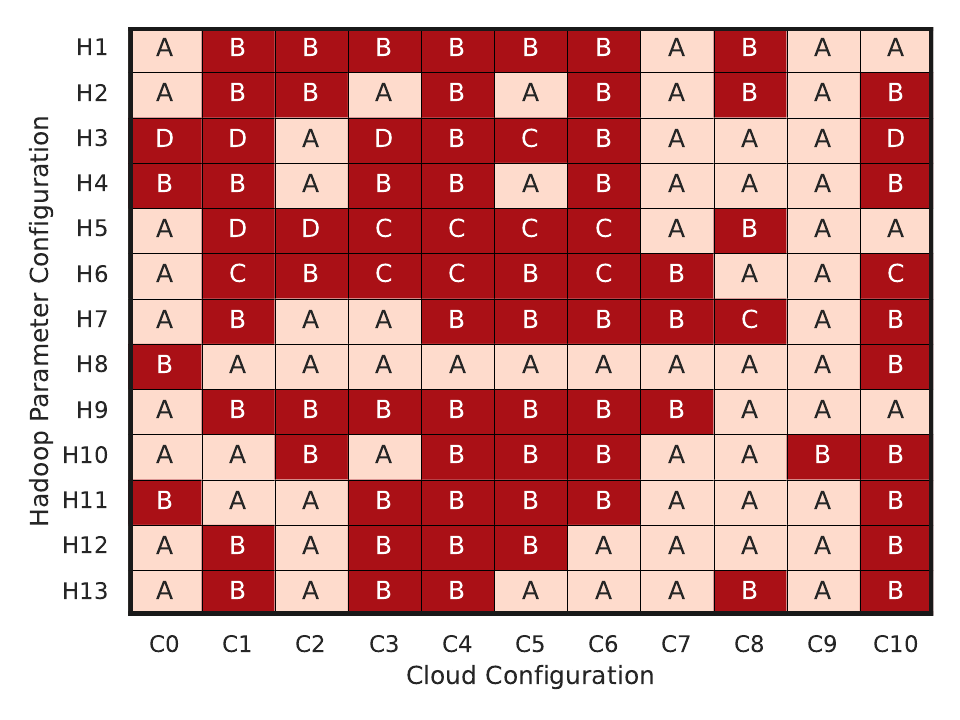}
  \caption{Word Count}
  \label{hh_wordcount_time}
\end{subfigure}
\begin{subfigure}{.32\textwidth}
  \centering
  \includegraphics[trim=10 10 10 10, clip, scale = 0.33]{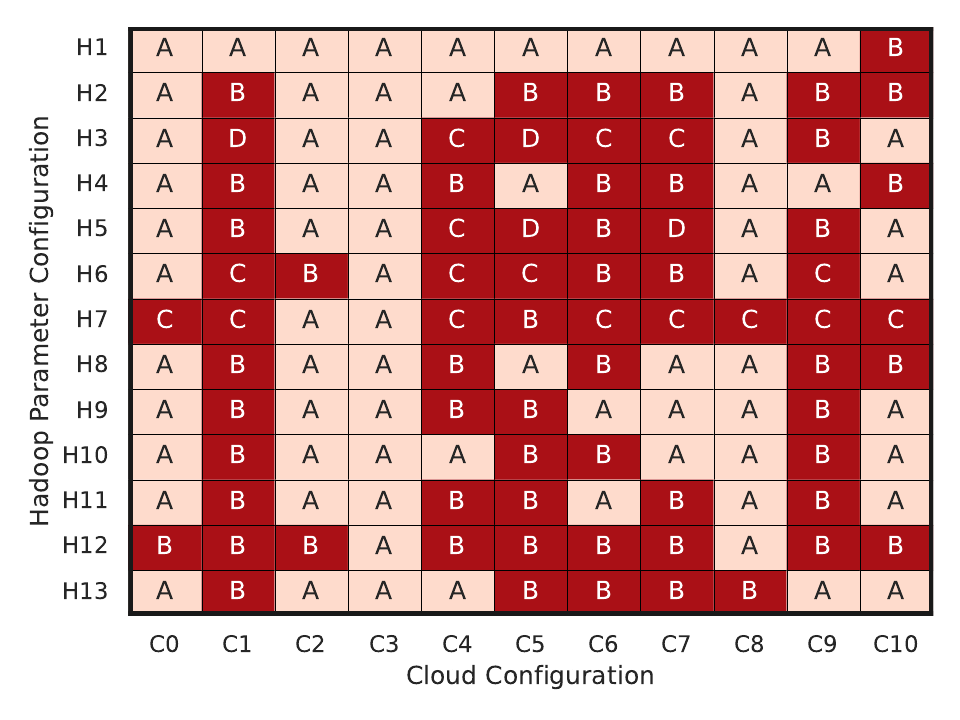}
  \caption{K-means}
  \label{hh_kmeans_time}
\end{subfigure}
\caption{Optimal values (A, B, C, and D) for \textbf{Hadoop} parameters with respect to various cloud configurations. ‘A’ denotes default value and ‘C’, ‘D’ and ‘E’ denote modified values of Hadoop parameters presented in Table \ref{hadoop_parameters_table}}
\label{hadoop_optimal_values}
\end{figure*}

Figure \ref{hadoop_execution_time} shows the execution time of Hadoop with respect to various cloud configurations and Hadoop parameter configurations. From Figure \ref{hadoop_execution_time}, we observe that the execution time varies both across columns (cloud configurations) as well as across rows (platform configurations). For example, as depicted in Figure \ref{hh_sort_time}, within the same row \textit{H(Def)}, the execution time varies from 76s to 168s. Similarly, keeping the column \textit{C0} fixed, the execution time varies from 70s to 115s across rows. Figure \ref{hadoop_optimal_values} shows the optimal value of each Hadoop parameter for each cloud configuration. By optimal value, we mean the value of Hadoop parameter that yields minimum execution time as reported in Figure \ref{hadoop_execution_time}. For instance, as depicted in Figure \ref{h_sort_time}, for cloud configuration \textit{C0}, platform configuration \textit{H3(B)} is better than \textit{H3(A)} (i.e., 70s vs. 79s). However, for cloud configuration \textit{C1}, platform configuration \textit{H3(A)} is better than \textit{H3(B)} (i.e., 108s vs. 129s). In Figure \ref{hadoop_optimal_values}, \textit{A} is the default value of the parameter while \textit{B}, \textit{C}, and \textit{D} are the modified values reported in Table \ref{hadoop_parameters_table}. Figure \ref{hadoop_optimal_values} reveals that as the cloud configuration changes, the optimal Hadoop configuration also changes. For example, optimal platform configuration for cloud configuration \textit{C0} is \textit{\{A, B, C, A, B, C, B, B, B, B, B, B, B\}}. On the other hand, optimal platform configuration for \textit{C1} is \textit{\{A, B, B, A, B, B, C, B, B, B, B, B, B\}}. In fact, there is not even one Hadoop configuration that is optimal for multiple cloud configurations. In other words, each time the cloud configuration changes so does the optimal configuration of Hadoop. Hence, we conclude that Hadoop parameter configuration is dependent upon cloud configuration.

\begin{figure*}
\captionsetup{justification=centering}
\centering
\begin{subfigure}{.32\textwidth}
  \centering
  \includegraphics[width=\linewidth]{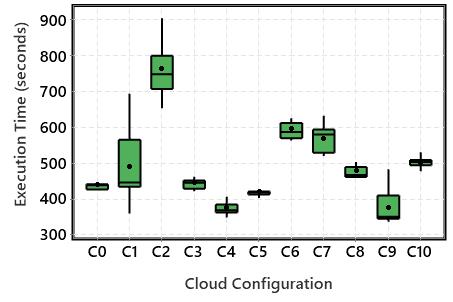}
  \caption{Impact of Hadoop configuration}
  %\vspace{5.00mm}
  \label{hadoop_boxplot_a}
\end{subfigure}%
\hspace{0.2 em}
\begin{subfigure}{.32\textwidth}
  \centering
  \includegraphics[width=\linewidth]{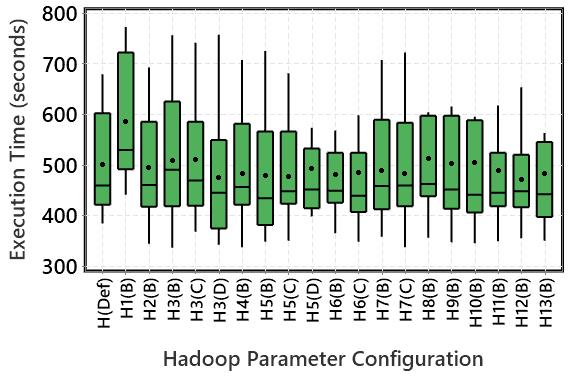}
  \caption{Impact of cloud configuration}
  %\vspace{5.00mm}
  \label{hadoop_boxplot_b}
\end{subfigure}%
\caption{Impact of cloud configuration and \textbf{Hadoop} configuration on execution time of K-means workload. \textit{C0}, \textit{C1}, and so on denote cloud configurations presented in Table \ref{cloud_configurations_table}. \textit{H(n)} denotes Hadoop parameter configurations reported in Table \ref{hadoop_parameters_table}. Each boxplot presents the values for execution time obtained with various Hadoop configurations in (a) and cloud configurations in (b) }
% In n\_m at x-axis, n and m denote the number of VMs deployed on the private and public cloud respectively
\label{hadoop_boxplot}
\hspace{2.0 em}
\end{figure*}

\begin{figure*}[t]
\centering
\begin{subfigure}{.32\textwidth}
  \centering
  \includegraphics[trim=0 0 15 10, clip, scale = 0.34]{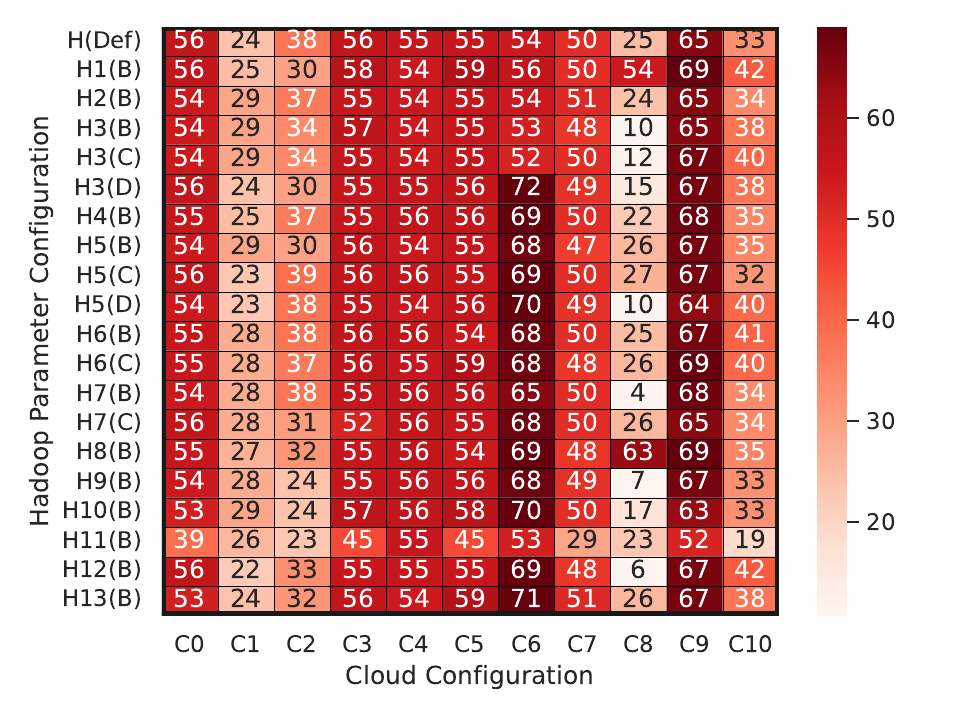}
  \caption{CPU Utilization (\%)}
  \label{h_cpu}
\end{subfigure}%
\begin{subfigure}{.32\textwidth}
  \centering
  \includegraphics[trim=0 0 15 10, clip, scale = 0.34]{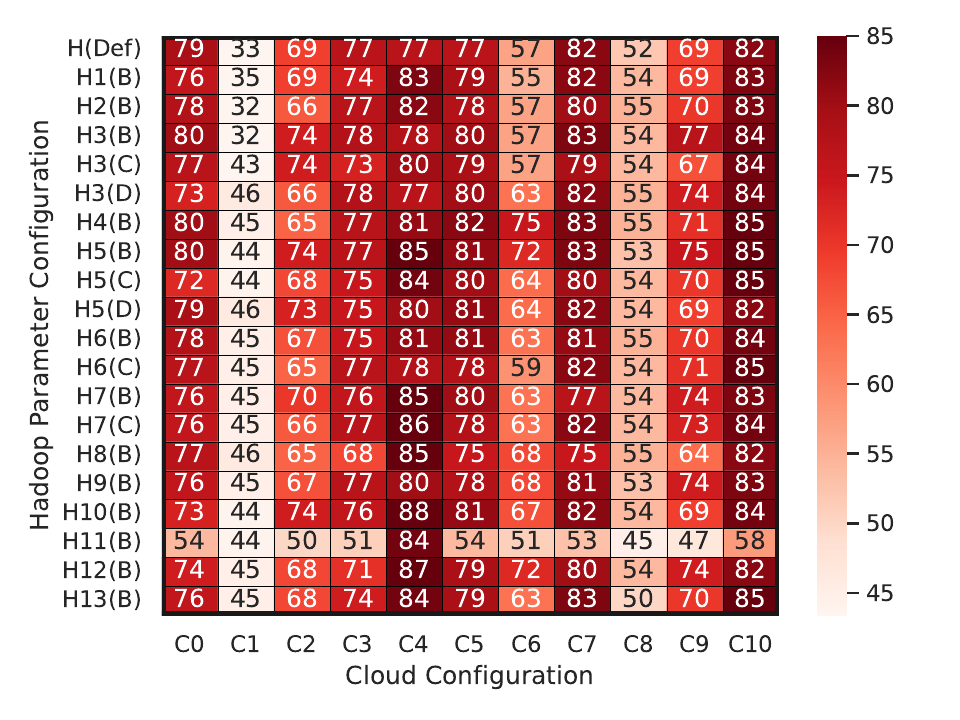}
  \caption{Memory Utilization (\%)}
  \label{h_memory}
\end{subfigure}
\begin{subfigure}{.32\textwidth}
  \centering
  \includegraphics[trim=0 0 15 10, clip, scale = 0.34]{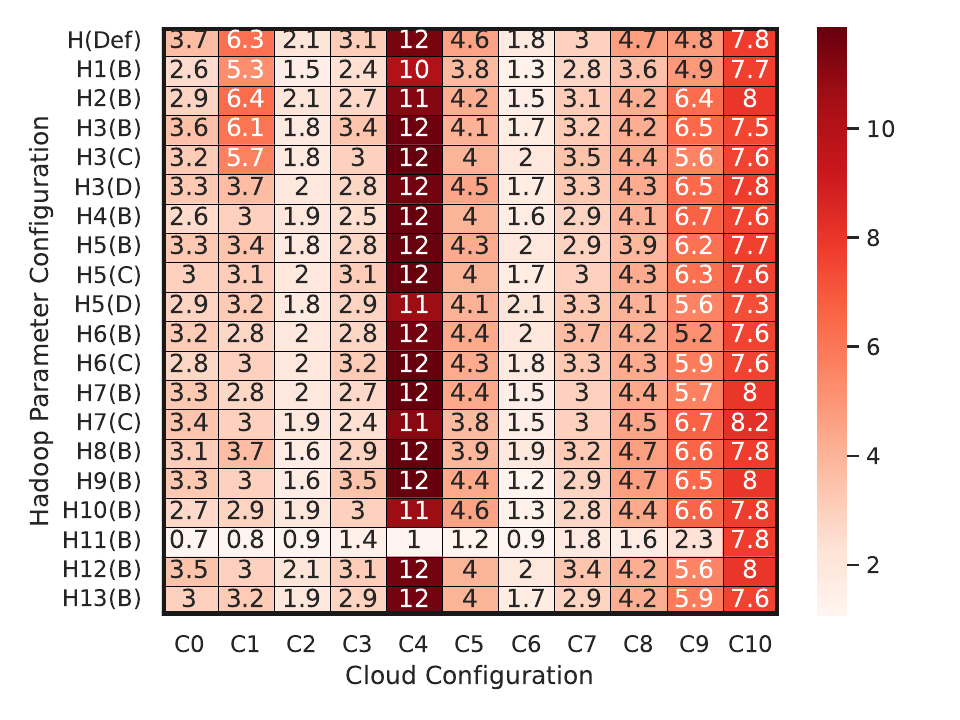}
  \caption{Hadoop - Disk Utilization (MB/s)}
  \label{h_network}
\end{subfigure}
\caption{Resource utilization of \textbf{Hadoop} with respect to various cloud configurations and Hadoop parameter configurations for executing \textbf{K-means}. \textit{C0}, \textit{C1}, and so on (x-axis) specify the cloud configurations presented in Table \ref{cloud_configurations_table}. \textit{H(n)} on y-axis denotes Hadoop parameter configurations reported in Table \ref{hadoop_parameters_table}}
\label{hadoop_resource_utilization}
\end{figure*}

We observe from Figure \ref{hadoop_execution_time} that the execution time varies more significantly as we move across cloud configurations (columns) as compared to the change in execution time across Hadoop configurations (rows). To elaborate on this point, we take the example of K-means and show the distribution of points (execution time) via boxplots in Figure \ref{hadoop_boxplot}. The more the height of the boxplots in Figure \ref{hadoop_boxplot}, the higher the variance in execution time as a consequence of the change in the configuration of cloud and/or Hadoop. It is evident that the height of the boxplots in Figure \ref{hadoop_boxplot_a} is lower than the height of the boxplots in Figure \ref{hadoop_boxplot_b}. This implies that the impact of cloud configuration on the execution time is higher than the impact of Hadoop configuration on the execution time. This is because cloud configuration has a more significant impact on resource utilization, especially CPU and memory, as shown in Figure \ref{hadoop_resource_utilization}. As we move across cloud configurations, the CPU and memory usage change drastically as compared to variation across rows, which results in impacting the execution time. We conclude that while configuration tuning of both cloud and Hadoop is important, it is crucial to pay more attention to cloud configuration as compared to Hadoop configuration. 

In Figure \ref{hadoop_execution_time}, as we move from \textit{C0} to \textit{C10} along x-axis, we observe a general decrease in the execution time as evident from the colours changing from red to white. As we move from \textit{C0} to \textit{C1}, the number of nodes in the cluster configuration increases (Table \ref{cloud_configurations_table}). For instance, 2 nodes are available in \textit{C0}, \textit{C1}, \textit{C2}, \textit{C3}, 3 nodes in \textit{C4}, \textit{C5}, \textit{C6}, 4 nodes in \textit{C7}, \textit{C8}, and 5 nodes in \textit{C9}, \textit{C10}.  This trend indicates that the execution time of Hadoop decreases as the number of nodes in the cluster increases. This finding is attributed to the parallelism model of Hadoop, which exploits horizontal scaling much better as compared to vertical scaling \cite{ullah2022evaluation}. In other words, Hadoop performs better with a cluster having more nodes of small size than a cluster having few nodes of larger size.

We also assess the impact of the homogeneous/heterogeneous nature of the cloud configuration on the execution time. By homogeneous cloud configuration, we mean a configuration (i.e., \textit{C0} and \textit{C9}) where all nodes are of the same type. On the other hand, heterogeneous cloud configuration (i.e., \textit{C1}, \textit{C2}, \textit{C3}, \textit{C4}, \textit{C5}, \textit{C6}, \textit{C7}, \textit{C8}, and \textit{C10}) has nodes of different types/flavours. Overall, we observe that the mean execution time with homogeneous cluster is lower than in heterogeneous clusters. Specifically, the mean execution time of Hadoop with heterogeneous clusters is 10.3\% higher than the execution time with homogeneous clusters. This is because of the load distribution model of Hadoop, which uses the YARN manager to distribute load among the workers. By default, YARN distributes the load equally among workers, which is quite favourable for homogeneous clusters as all nodes are of the same flavour. However, such equal distribution is not efficient for heterogeneous clusters where some nodes have more resources as compared to others. In such a case, the nodes with low resources become a bottleneck for the nodes with higher resources. This is because all nodes in the cluster work collaboratively, therefore, nodes with more resources have to wait for the nodes with lower resources to complete their task before proceeding on to the next intermediate task.

With respect to workloads, the index of variation for the values presented in Figure \ref{h_sort_time}, Figure \ref{h_wordcount_time}, and Figure \ref{h_kmeans_time} are 0.34, 0.44, and 0.45, respectively. Such variation is also evident from the range of values i.e., 68s - 281s for Sort, 105s - 390s for Word Count, and 337s - 949s for K-means presented in Figure \ref{hadoop_execution_time}. This indicates that the impact of cloud and Hadoop configuration is more significant for iterative workloads as compared to batch workloads. A potential reason is that iterative workloads are computationally more expensive as compared to batch workloads. For instance, as per our results, the mean execution time of Hadoop is 105s, 198s, and 496s for Sort, Word Count, and K-means, respectively. Since iterative workloads are computationally expensive, the pain and gain of configuration tuning are higher for iterative workloads.

\begin{figure*}[t]
\centering
\begin{subfigure}{.32\textwidth}
  \centering
  \includegraphics[trim=0 0 15 10, clip, scale = 0.34]{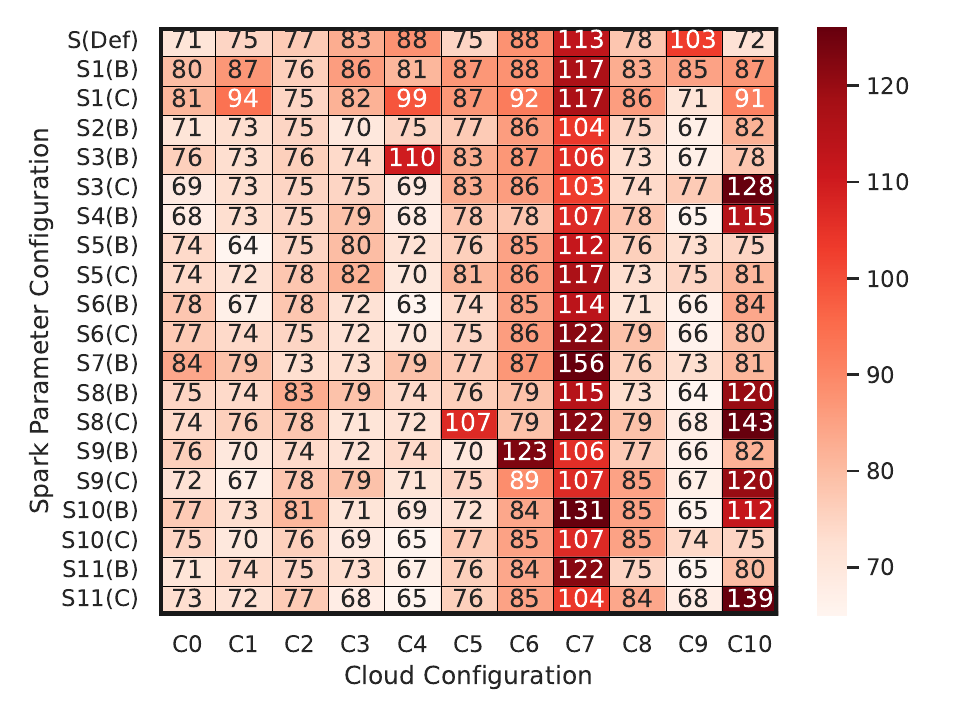}
  \caption{Sort}
  \label{s_sort}
\end{subfigure}%
\begin{subfigure}{.32\textwidth}
  \centering
  \includegraphics[trim=0 0 15 10, clip, scale = 0.34]{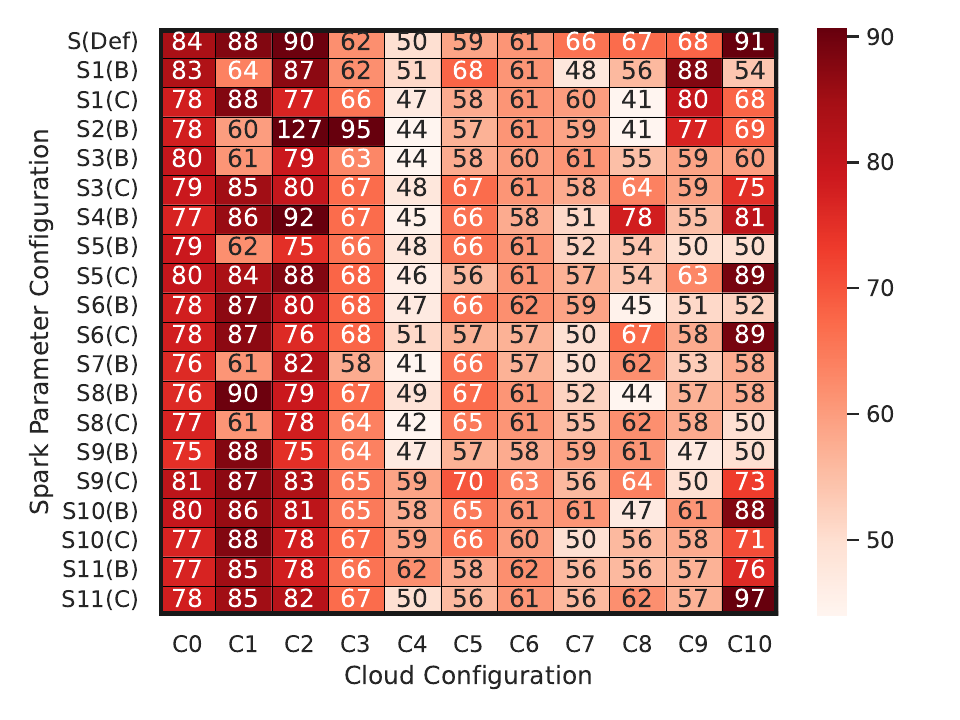}
  \caption{Word Count}
  \label{s_wordcount}
\end{subfigure}
\begin{subfigure}{.32\textwidth}
  \centering
  \includegraphics[trim=0 0 15 10, clip, scale = 0.34]{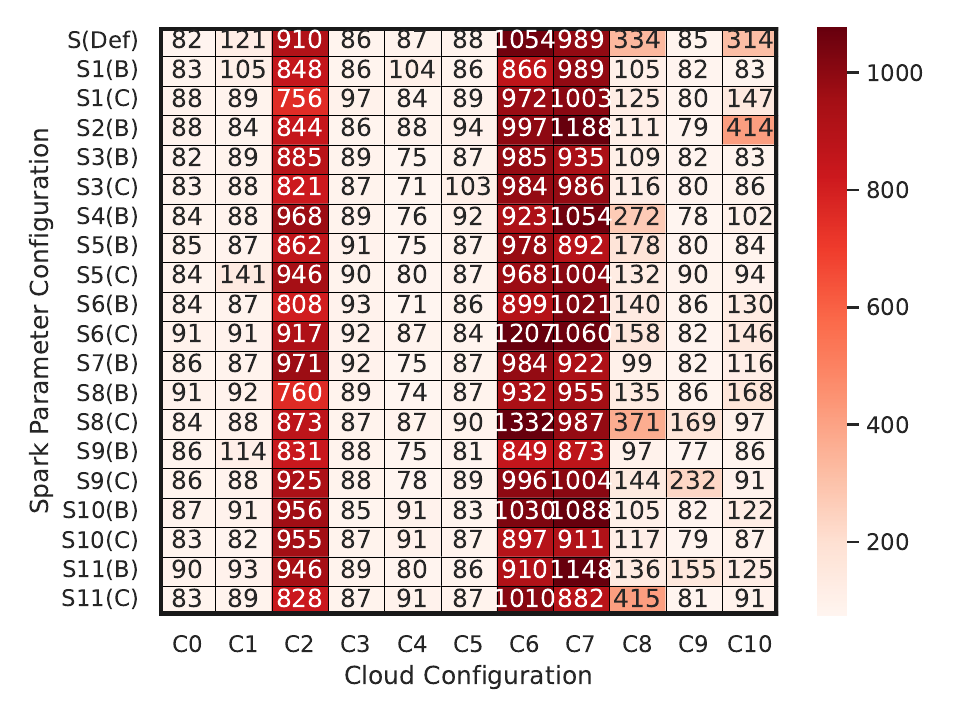}
  \caption{K-means}
  \label{s_kmeans}
\end{subfigure}
\caption{Execution time of \textbf{Spark} with respect to various cloud configurations and Spark parameter configurations for executing Sort, Word Count, and K-means. \textit{C0}, \textit{C1}, and so on (x-axis) specify the cloud configurations presented in Table \ref{cloud_configurations_table}. \textit{S(n)} on y-axis denotes Spark parameter configurations reported in Table \ref{spark_parameters_table}}
\label{spark_execution_time}
\end{figure*}

\begin{figure*}[t]
\centering
\begin{subfigure}{.32\textwidth}
  \centering
  \includegraphics[trim=10 10 10 10, clip, scale = 0.33]{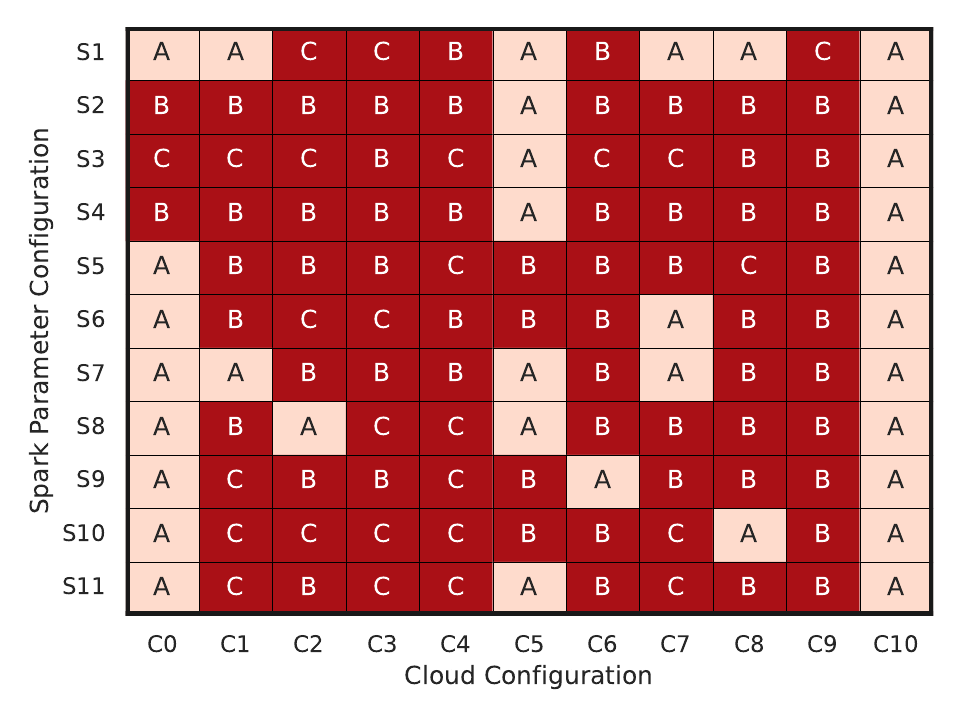}
  \caption{Sort}
  \label{s_sort_opt}
\end{subfigure}%
\begin{subfigure}{.32\textwidth}
  \centering
  \includegraphics[trim=10 10 10 10, clip, scale = 0.33]{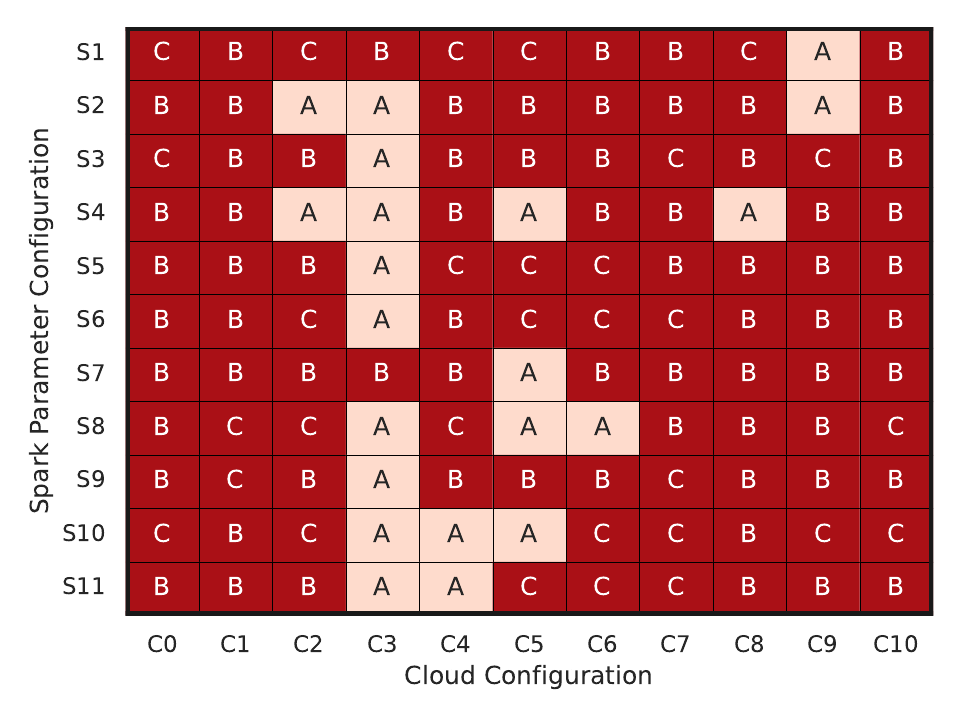}
  \caption{Word Count}
  \label{s_wordcount_opt}
\end{subfigure}
\begin{subfigure}{.32\textwidth}
  \centering
  \includegraphics[trim=10 10 10 10, clip, scale = 0.33]{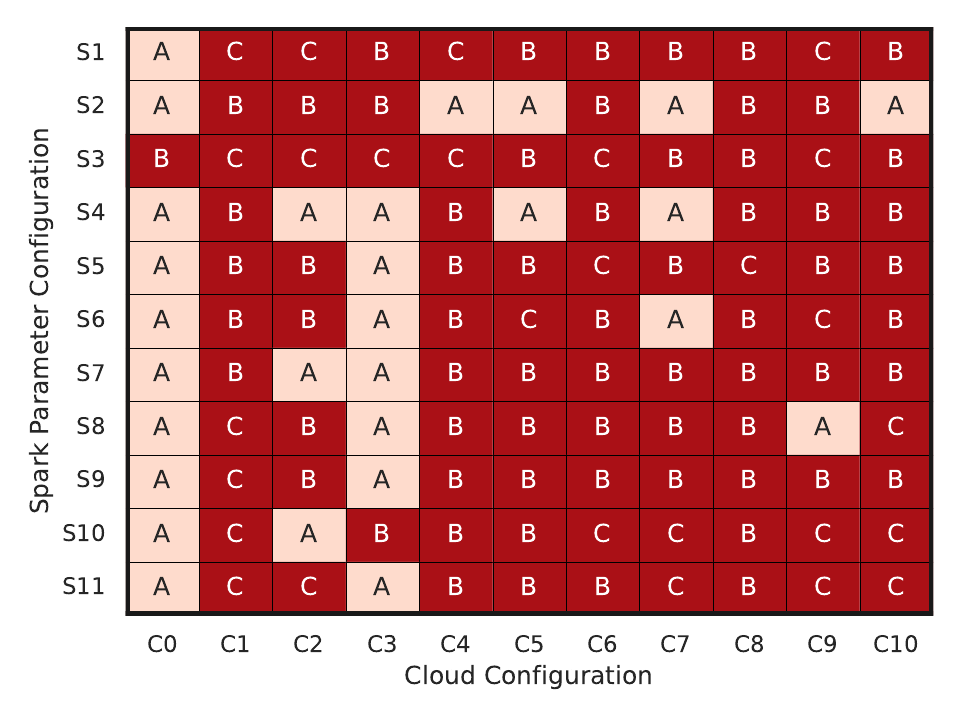}
  \caption{K-means}
  \label{s_kmeans_opt}
\end{subfigure}
\caption{Optimal values (A, B, and C) for \textbf{Spark} parameters with respect to various cloud configurations. 'A' denotes default value and 'B' and 'C' denote modified values of Spark parameters presented in Table \ref{spark_parameters_table}}
\label{spark_optimal_values}
\end{figure*}

\begin{figure*}[t]
\captionsetup{justification=centering}
\centering
\begin{subfigure}{.32\textwidth}
  \centering
  \includegraphics[width=\linewidth]{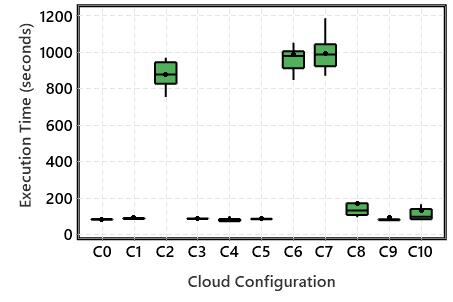}
  \caption{Impact of Spark configuration}
  %\vspace{5.00mm}
  \label{spark_boxplot_a}
\end{subfigure}%
\hspace{0.2 em}
\begin{subfigure}{.32\textwidth}
  \centering
  \includegraphics[width=\linewidth]{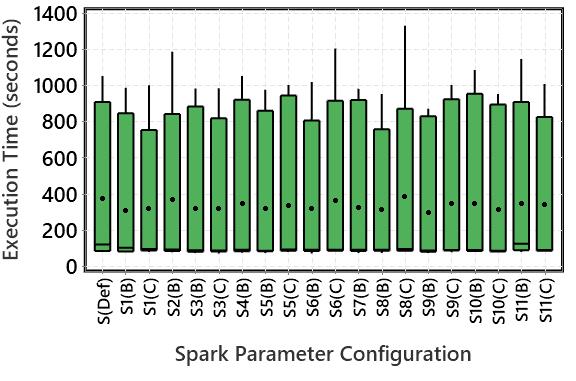}
  \caption{Impact of cloud configuration}
  %\vspace{5.00mm}
  \label{spark_boxplot_b}
\end{subfigure}%
\caption{Impact of cloud configuration and \textbf{Spark} configuration on the execution time of K-means workload. \textit{C0}, \textit{C1}, and so on denote cloud configurations presented in Table \ref{cloud_configurations_table}. \textit{S(n)} denotes Spark parameter configurations reported in Table \ref{spark_parameters_table}. Each boxplot presents the values for execution time obtained with various Spark configurations in (a) and cloud configurations in (b) }
% In n\_m at x-axis, n and m denote the number of VMs deployed on the private and public cloud respectively
\label{spark_boxplot}
\hspace{2.0 em}
\end{figure*}

\begin{figure*}[t]
\centering
\begin{subfigure}{.32\textwidth}
  \centering
  \includegraphics[trim=0 0 15 10, clip, scale = 0.34]{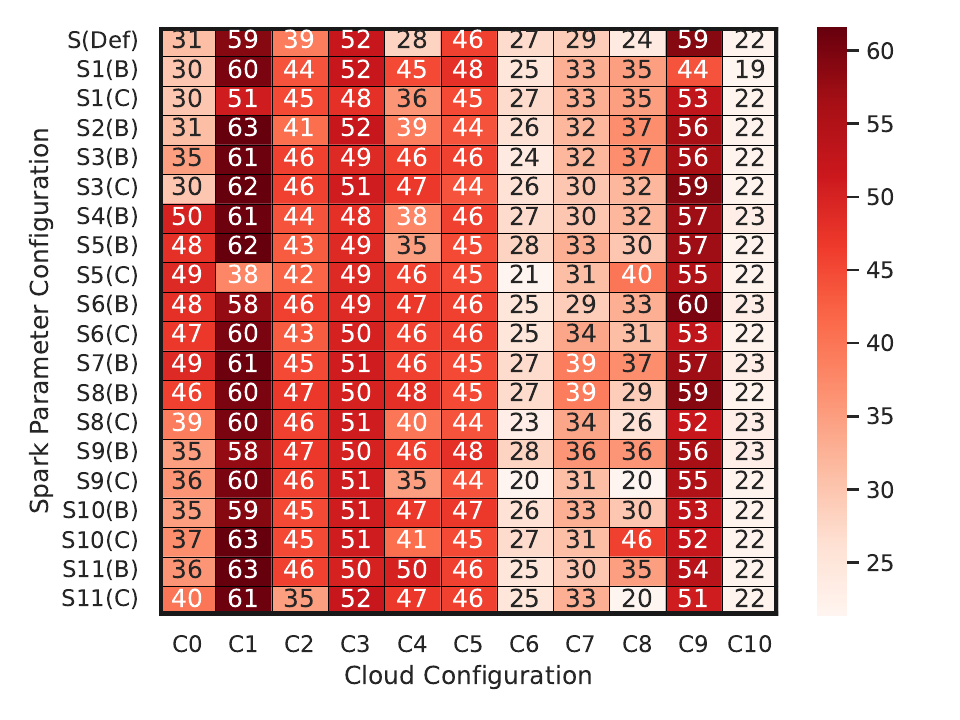}
  \caption{CPU Utilization (\%)}
  \label{s_cpu}
\end{subfigure}%
\begin{subfigure}{.32\textwidth}
  \centering
  \includegraphics[trim=0 0 15 10, clip, scale = 0.34]{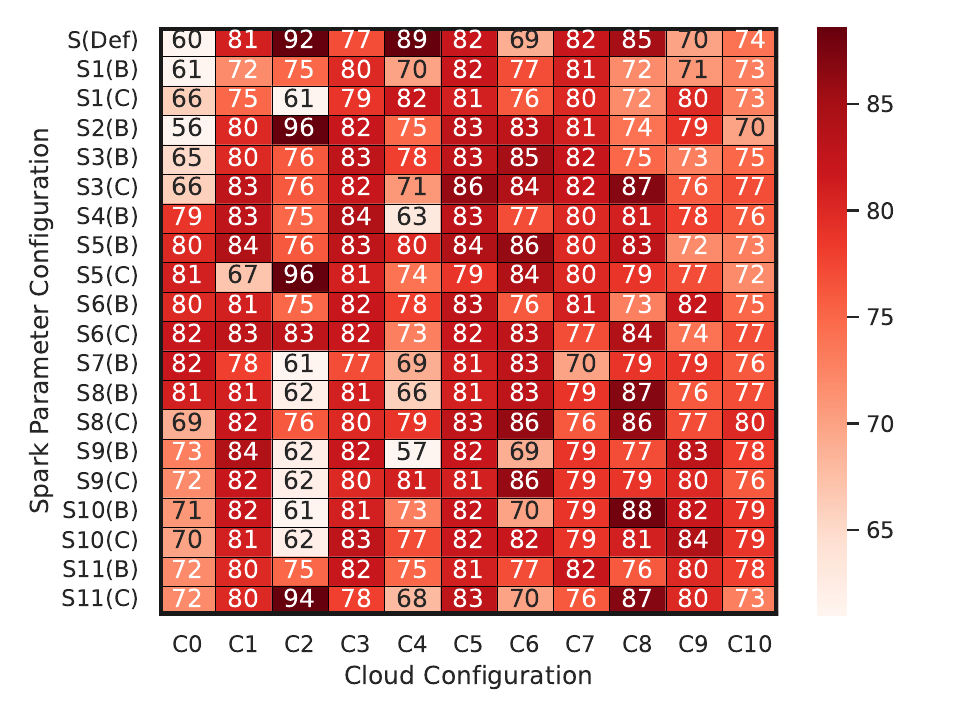}
  \caption{Spark - Memory Utilization (\%)}
  \label{s_memory}
\end{subfigure}
\begin{subfigure}{.32\textwidth}
  \centering
  \includegraphics[trim=0 0 15 10, clip, scale = 0.34]{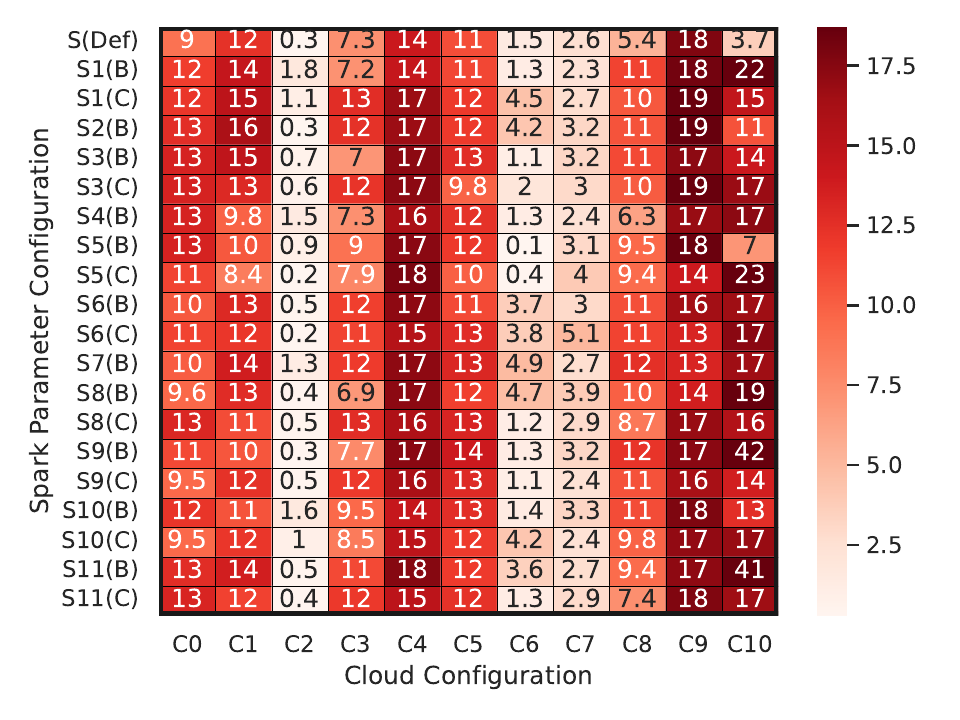}
  \caption{Spark - Disk Utilization (MB/s)}
  \label{s_disk}
\end{subfigure}
\caption{Resource utilization of \textbf{Spark} with respect to various cloud configurations and Spark parameter configurations for executing \textbf{K-means}. \textit{C0}, \textit{C1}, and so on (x-axis) specify the cloud configurations presented in Table \ref{cloud_configurations_table}. \textit{S(n)} on y-axis denotes Spark parameter configurations reported in Table \ref{spark_parameters_table}}
\label{spark_resource_utilization}
\end{figure*}

\subsection{Impact of Cloud Configuration on Spark Configuration}

Figure \ref{spark_execution_time} shows the execution time of Spark with respect to various cloud configurations and Spark configurations for the three workloads. Similar to Hadoop, the execution time varies both across cloud configurations as well as across Spark parameter configurations. As shown in Figure \ref{spark_optimal_values}, the optimal values of Spark parameters varies with respect to cloud configurations. As an example, for Sort workload, cloud configurations \textit{C0, C1, C5, C7, and C8} require Spark parameter \textit{S1} to be tuned to default value \textit{A}. On the other hand, cloud configuration \textit{C6} requires \textit{S1} be tuned to \textit{B} and \textit{C2, C3, and C9} require it to tuned to \textit{C} for Sort workload. In all 33 cases (11 cloud configurations $\times$ 3 workloads) for Spark presented in Figure \ref{spark_optimal_values}, the optimal configurations of Spark are different from each other. This indicates that optimal configuration of Spark varies with respect to the underlying cloud configuration. Hence, we conclude that cloud configuration impacts Spark configuration. From Figure \ref{spark_optimal_values}, we also observe that irrespective of the underlying cloud configuration, the default values of Spark parameters are mostly (74.9\%) not optimal. This can be observed from red color highlighting modified values (\textit{B and C}) that yield lower execution time as compared to default value \textit{A}.

We also assess whether or not the impact of cloud configuration on execution time is more significant than the impact of Spark configuration on execution time. In this regard, Figure \ref{spark_execution_time} reveals that the variation in heat as we move from one column to another is higher than the variation in heat when we move from one row to another. To have a more clearer view, we present the distribution of execution time for K-means with respect to changing Spark parameter configuration and with respect to changing cloud configuration in Figure \ref{spark_boxplot_a} and Figure \ref{spark_boxplot_b}, respectively. This means that in Figure \ref{spark_boxplot_a}, the cloud configuration is fixed and only the Spark parameter configuration changes. On the other hand, in Figure \ref{spark_boxplot_b}, Spark configuration is set to default and cloud configuration changes. It is evident that the distribution of points in boxplots in Figure \ref{spark_boxplot_b} is higher than the distribution of points in boxplots in Figure \ref{spark_boxplot_a}. This implies that the configuration of cloud has a more significant impact on execution time as compared to the configuration of Spark parameters. 

Although not as significant as cloud configuration, we observe variation in execution time as Spark parameter configuration changes. For instance, our results reveal that the optimal tuning of Spark parameter \textit{S6} can reduce execution time by up to 78\%. Furthermore, the optimal value of a parameter is different for different workloads. For example, as per our results, the optimal value of Spark parameter \textit{S3} with cloud configuration \textit{C3} is \textit{‘C’} for Sort, \textit{‘A’} for Word Count, and \textit{‘B’} for K-means. This further strengthens our assertion that in addition to a cloud configuration, Spark parameter configuration is also crucial to maximize the performance of Spark. Unlike Hadoop, the impact of the increase/decrease in the number of VMs is not so evident for Spark. As an example, for all three workloads, mostly cloud configuration \textit{C4} and \textit{C5} yield lower execution time with some variation depending upon Spark configuration. This is because with cloud configuration \textit{C4} and \textit{C5}, the resources like CPU and RAM are heavily utilized as shown in Figure \ref{spark_resource_utilization}. On the other hand, a very low number of VMs (e.g., in cloud configuration \textit{C0}) or high number of VMs (e.g., in cloud configuration \textit{C10}) leads to a comparatively higher execution time. Therefore, unlike Hadoop that favours a large number of VMs of small size, Spark favours a medium number of VMs of medium size. With the exception of \textit{C0} for Word Count, Figure \ref{spark_execution_time} shows that the execution time obtained with homogeneous cluster (i.e. \textit{C0} and \textit{C9}) is lower than than the execution time obtained with homogeneous cluster. This is attributed to the resource manager YARN used for Spark, which allocates load equally among the workers. This way, the workers with lower resources in the cluster become a bottleneck for the nodes with higher resources. Consequently, the heterogeneous cluster is not utilized to the maximum of its potential.

\begin{figure*}[t]
\centering
\begin{subfigure}{.32\textwidth}
  \centering
  \includegraphics[trim=0 0 15 10, clip, scale = 0.34]{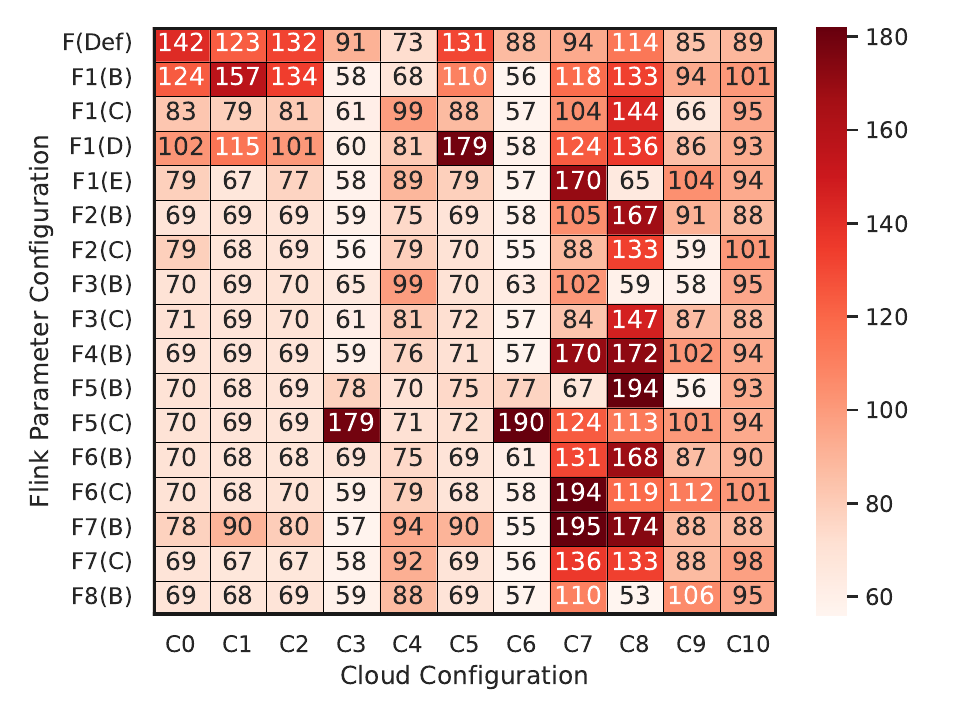}
  \caption{Sort}
  \label{f_sort}
\end{subfigure}%
\begin{subfigure}{.32\textwidth}
  \centering
 \includegraphics[trim=0 0 15 10, clip, scale = 0.34]{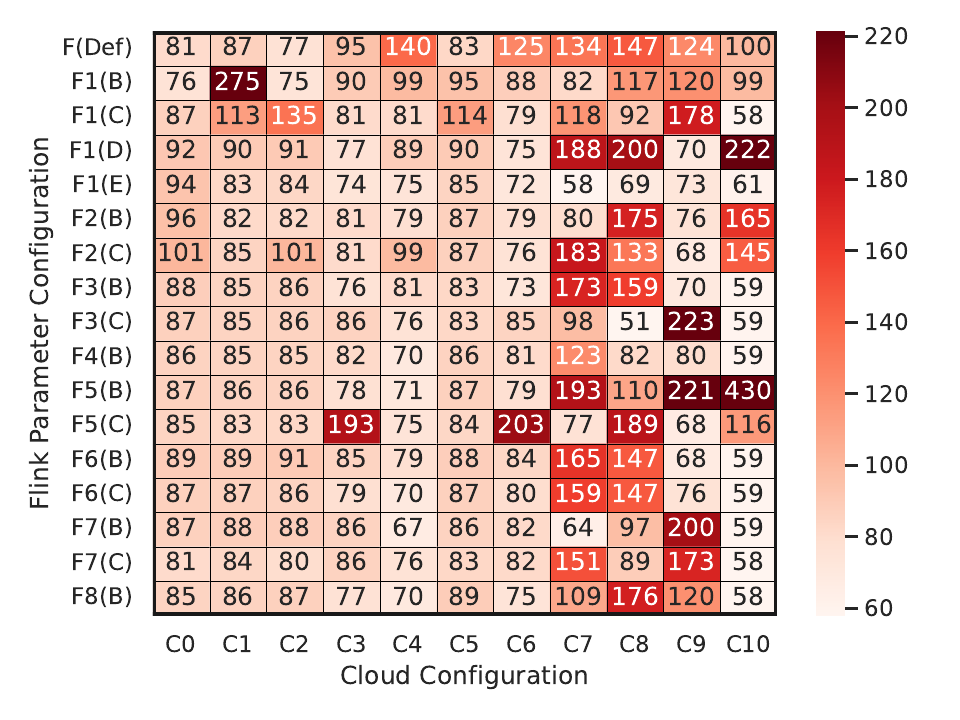}
  \caption{Word Count}
  \label{f_wordcount}
\end{subfigure}
\begin{subfigure}{.32\textwidth}
  \centering
  \includegraphics[trim=0 0 15 10, clip, scale = 0.34]{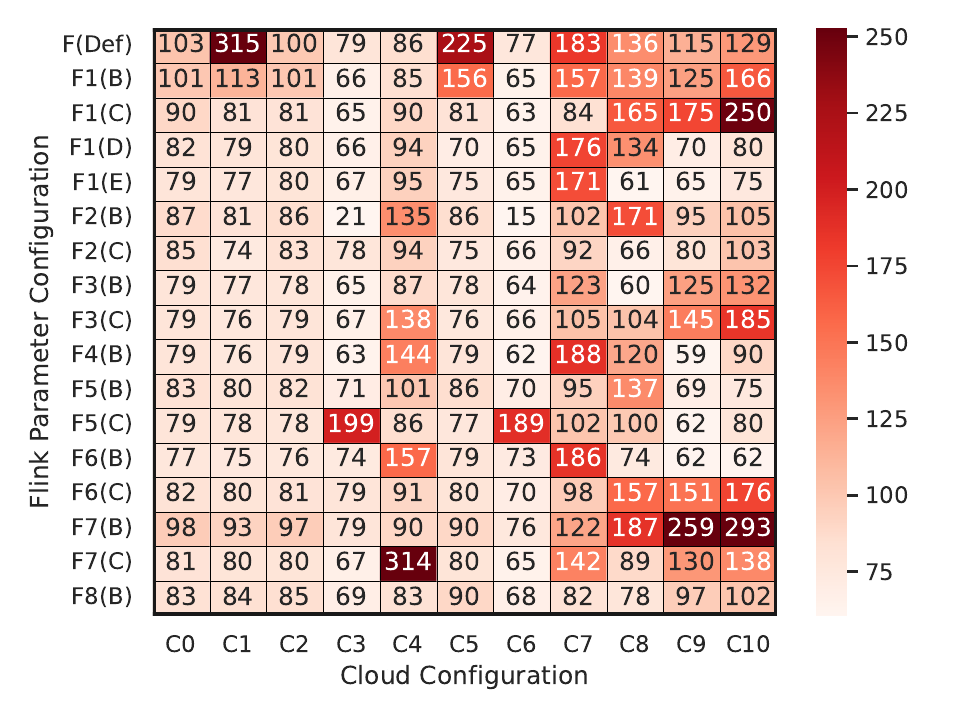}
  \caption{K-means}
  \label{f_kmeans}
\end{subfigure}
\caption{Execution time of \textbf{Flink} with respect to various cloud configurations and Flink parameter configurations for executing Sort, Word Count, and K-means. \textit{C0}, \textit{C1}, and so on (x-axis) specify the cloud configurations presented in Table \ref{cloud_configurations_table}. \textit{F(n)} on y-axis denotes Flink parameter configurations reported in Table \ref{flink_parameters_table}}
\label{flink_execution_time}
\end{figure*}

\begin{figure*}[t]
\centering
\begin{subfigure}{.32\textwidth}
  \centering
  \includegraphics[trim=10 10 10 10, clip, scale = 0.33]{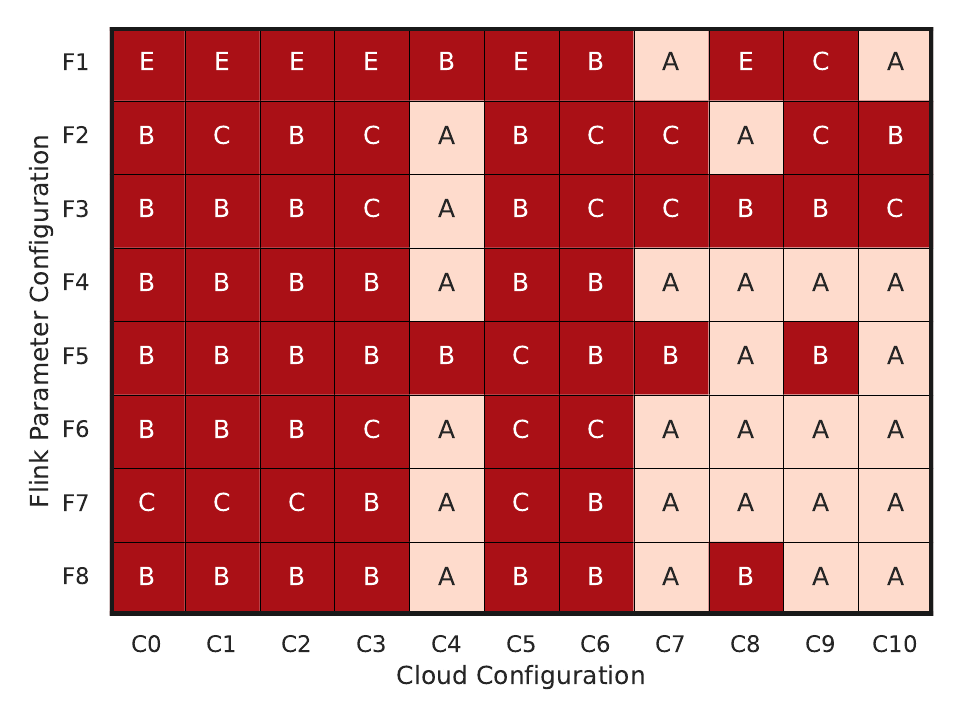}
  \caption{Sort}
  \label{f_sort_opt}
\end{subfigure}%
\begin{subfigure}{.32\textwidth}
  \centering
  \includegraphics[trim=10 10 10 10, clip, scale = 0.33]{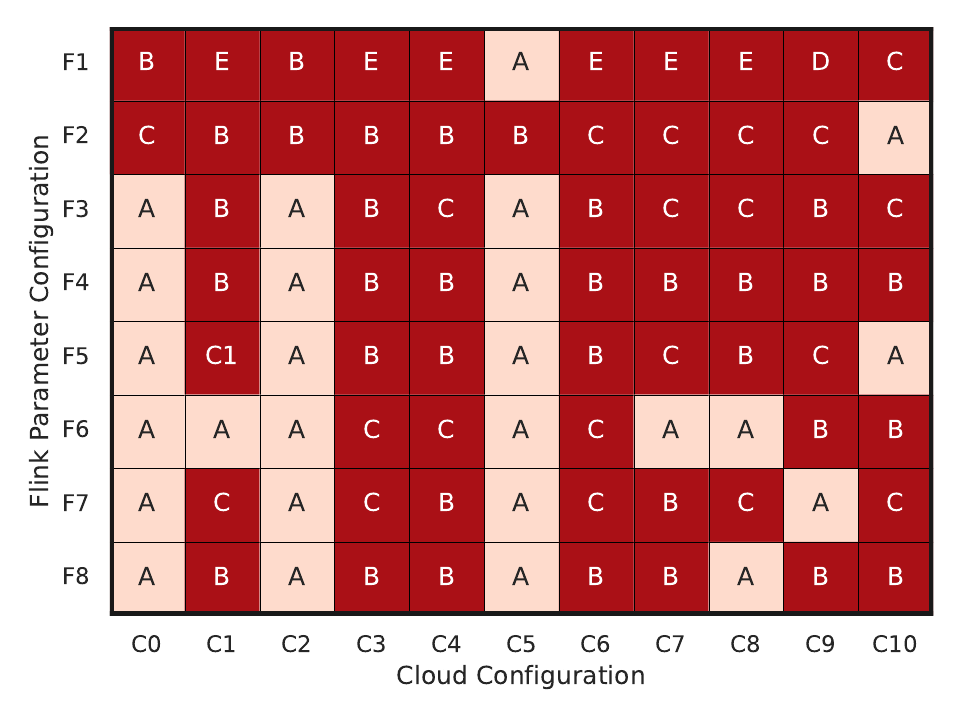}
  \caption{Word Count}
  \label{f_wordcount_opt}
\end{subfigure}
\begin{subfigure}{.32\textwidth}
  \centering
  \includegraphics[trim=10 10 10 10, clip, scale = 0.33]{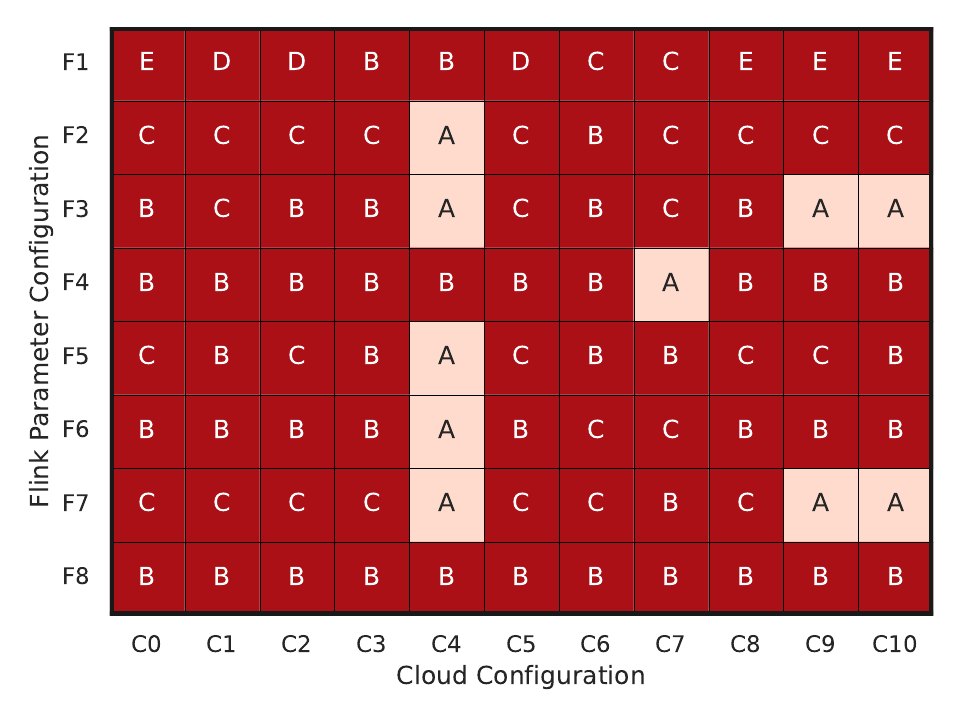}
  \caption{K-means}
  \label{f_kmeans_opt}
\end{subfigure}
\caption{Optimal values (A, B, C, D, and E) for \textbf{Flink} parameters with respect to various cloud configurations. ‘A’ denotes default value and ‘B’, ‘C’, ‘D’, and ‘E’ denote modified values presented in Table \ref{flink_parameters_table}}
\label{flink_optimal_values}
\end{figure*}

\begin{figure*}[t]
\captionsetup{justification=centering}
\centering
\begin{subfigure}{.32\textwidth}
  \centering
  \includegraphics[width=\linewidth]{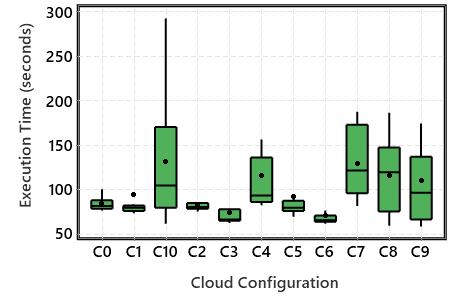}
  \caption{Impact of Flink configuration}
  %\vspace{5.00mm}
  \label{flink_boxplot_a}
\end{subfigure}%
\hspace{0.2 em}
\begin{subfigure}{.32\textwidth}
  \centering
  \includegraphics[width=\linewidth]{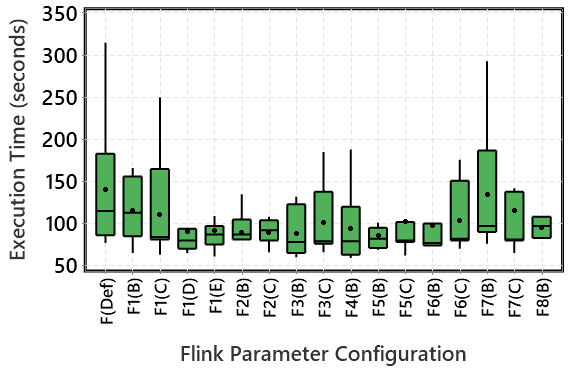}
  \caption{Impact of cloud configuration}
  %\vspace{5.00mm}
  \label{flink_boxplot_b}
\end{subfigure}%
\caption{Impact of cloud configuration and \textbf{Flink} configuration on the execution time of K-means workload. \textit{C0}, \textit{C1}, and so on denote cloud configurations presented in Table \ref{cloud_configurations_table}. \textit{F(n)} denotes Flink parameter configurations reported in Table \ref{flink_parameters_table}. Each boxplot presents the values for execution time obtained with various Flink configurations in (a) and cloud configurations in (b) }
% In n\_m at x-axis, n and m denote the number of VMs deployed on the private and public cloud respectively
\label{flink_boxplot}
\hspace{2.0 em}
\end{figure*}

\begin{figure*}[!tbp]
\centering
\begin{subfigure}{.32\textwidth}
  \centering
  \includegraphics[trim=0 0 15 10, clip, scale = 0.34]{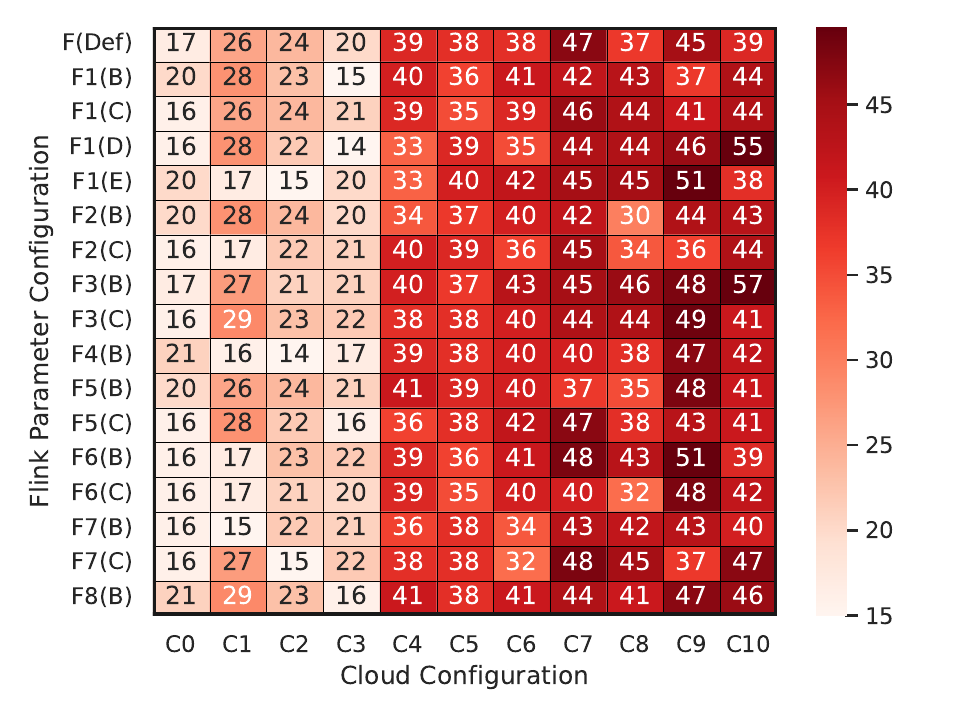}
  \caption{CPU Utilization (\%)}
  \label{f_cpu}
\end{subfigure}%
\begin{subfigure}{.32\textwidth}
  \centering
  \includegraphics[trim=0 0 15 10, clip, scale = 0.34]{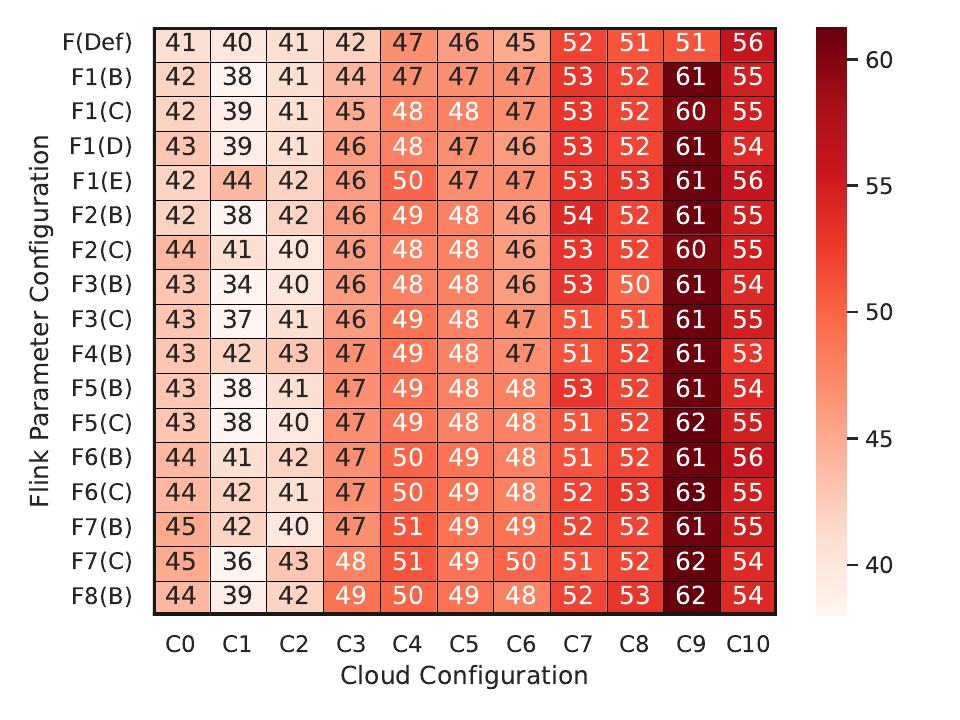}
  \caption{Memory Utilization (\%)}
  \label{f_memory}
\end{subfigure}
\begin{subfigure}{.32\textwidth}
  \centering
  \includegraphics[trim=0 0 15 10, clip, scale = 0.34]{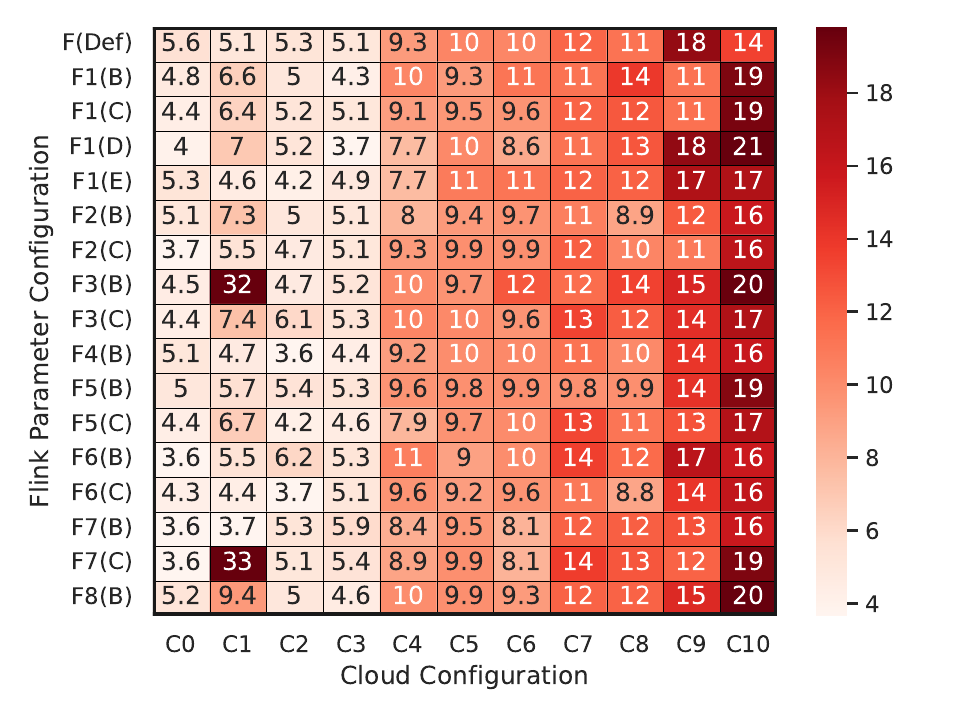}
  \caption{Disk Utilization (MB/s)}
  \label{f_disk}
\end{subfigure}
\caption{Resource utilization of \textbf{Flink} with respect to various cloud configurations and Flink parameter configurations for executing \textbf{K-means}. \textit{C0}, \textit{C1}, and so on (x-axis) specify the cloud configurations presented in Table \ref{cloud_configurations_table}. \textit{F(n)} on y-axis denotes Flink parameter configurations reported in Table \ref{flink_parameters_table}}
\label{flink_resource_utilization}
\end{figure*}

\subsection{Impact of Cloud Configuration on Flink Configuration}

Figure \ref{flink_execution_time} presents the execution time of Flink with respect to various cloud configurations and Flink configurations for Sort, Word Count, and K-means. Whilst we observe that the execution time changes with the change in the configuration of cloud and Flink, the change is not as evident as it was for Hadoop and Spark. A potential reason for this is the in-built optimizer of Flink, which is missing in Hadoop \cite{carbone2015apache}. This optimizer helps Flink to optimize itself in accordance with the operating environment that also underpins the cloud configuration. Moreover, unlike Flink, Hadoop and Spark do not have their own mechanism for automatic memory management, which could help them in garbage collection and avoiding memory leaks/spills to disk due to shortage of memory. Comparing Figure \ref{hadoop_execution_time}, Figure \ref{spark_execution_time}, and Figure \ref{flink_execution_time}, we observe that the execution time of Flink is generally lower than the execution time of Hadoop and Spark. Specifically, as per our results, the mean execution time of Flink is 67\% lower than Spark and 177\% lower than Hadoop. The main contributor to better Flink's performance is its better use of resources as shown in Figure \ref{flink_resource_utilization}, which indicates that the mean usage of CPU and RAM by Flink is around 34\% and 49\% of the total CPU and RAM capacity. 

Figure \ref{flink_optimal_values} depicts that similar to Hadoop and Spark, the optimal values of Flink parameters change with respect to cloud configurations. Each time cloud configuration changes, the optimal values of Flink configuration parameters also change. This guides us to conclude that cloud configuration impacts Flink configuration. In 76.9\% of the cases, default values are not optimal as shown in Figure \ref{flink_optimal_values} meaning that a user needs to manually/semi-automatically need to tune the values of various Flink parameters. Although the parameter values need to be tuned to optimal values, the gain from the Flink parameters is not as much as the gain from tuning Hadoop and Spark parameters. For instance, as per our results, optimal tuning of Flink reduces mean execution time by 14.7\% as compared to default Flink settings. In contrast, Hadoop and Spark parameter tuning reduce mean execution time by 43.6\% and 20.1\%, respectively, as compared to their default settings. 

With Flink, the difference between the execution time for the batch workload (Sort and Word Count) and iterative workload (K-means) is not as high as with Spark and especially Hadoop. This is because Flink is specifically designed for iterative workloads that are executed in a micro-stream fashion. Similar to Hadoop and Spark, we observe from Figure \ref{flink_boxplot} that the execution time for homogeneous cluster (i.e., \textit{C0} and \textit{C9}) is lower as compared to heterogeneous clusters. Figure \ref{flink_boxplot} depicts the impact of cloud and Flink configuration by showing the execution time values obtained for a fixed cloud or Flink configuration. It is evident that the points are not as distributed as was for Hadoop (Figure \ref{hadoop_boxplot}) and Spark (Figure \ref{spark_boxplot}). We calculated the variance among the points shown via boxplots in Figure \ref{flink_boxplot}. We found that the mean variance among the values for execution time shown in Figure \ref{flink_execution_time} is 36.6. These numbers were 39.2 and 54.1 for Hadoop and Spark, respectively. This implies that the impact of configuration tuning of Flink on the execution time is not as significant as the impact of configuration tuning on Hadoop and Spark. Finally, Figure \ref{flink_execution_time} shows that as we move along x-axis from \textit{C0} to \textit{C10}, the execution time generally increases. Since as we move from \textit{C0} to \textit{C10}, the number of nodes in the cluster increases, we can deduce that unlike Hadoop, Flink yields lower execution time with fewer nodes of large size as compared to more nodes of small size.

\section{Our Co-Tuning Approach}

\begin{figure*}
\centering
\begin{subfigure}{.30\textwidth}
  \centering
  \includegraphics[width=\linewidth]{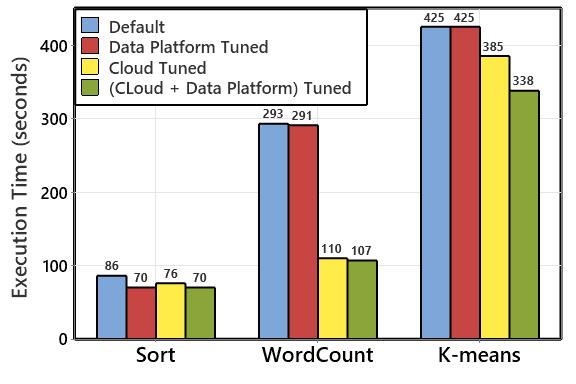}
  \caption{Hadoop}
  \label{hadoop-config-benefit}
\end{subfigure}%
\begin{subfigure}{.30\textwidth}
  \centering
  \includegraphics[width=\linewidth]{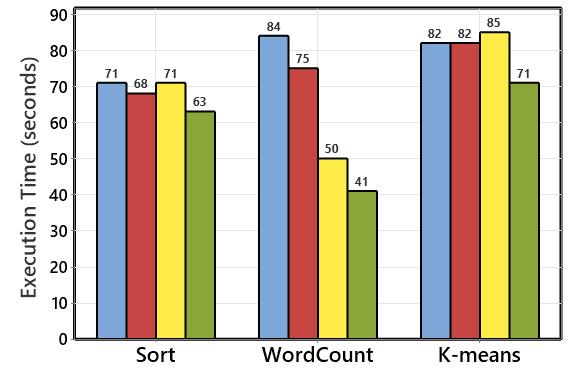}
  \caption{Spark}
  \label{spark-config-benefit}
\end{subfigure}
\begin{subfigure}{.30\textwidth}
  \centering
  \includegraphics[width=\linewidth]{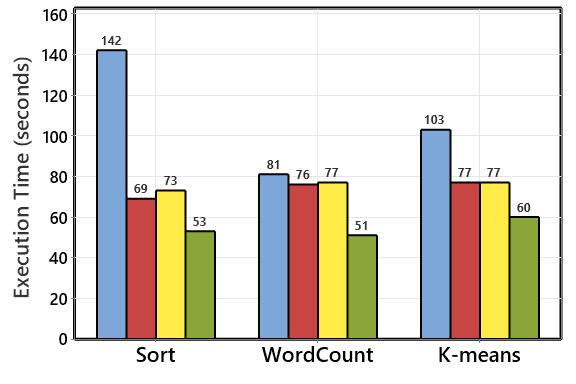}
  \caption{Flink}
  \label{flink-config-benefit}
\end{subfigure}
\caption{Comparison of execution time for four settings i.e., default settings, data platform tuned, cloud tuned, and (cloud + data platform) tuned of a distributed data processing system}
\label{tuning_benefit}
\end{figure*}

The results presented in the previous section revealed that cloud configuration impacts data platform configuration. Hence, it is important to co-tune cloud and data platforms to utilize the full potential of both cloud and data platforms. For example, users need to configure data platforms differently to achieve maximum performance from them when hosted in cloud setups with equal resources but varying configurations. To illustrate this, we present in Figure \ref{tuning_benefit} the maximum gain in execution time possible via tuning only the data platform, tuning only the cloud, and tuning both the data platform and cloud. These values are based on the data presented in Figure \ref{hadoop_execution_time}, Figure \ref{spark_execution_time}, and Figure \ref{flink_execution_time}. It is evident from Figure \ref{tuning_benefit} that almost in every case tuning reduces execution time as compared to the execution time with default settings. Furthermore, we observe that on average, tuning data platform, cloud, and (data platform + cloud) can reduce execution time up to 12.9\%, 22.4\%, and 35.4\% as compared to execution time with default settings. This means that in comparison to only tuning the data platform or only tuning the cloud, tuning both the data platform and cloud at the same time reduces the execution time more significantly. Hence, it is important to co-tune both the data platform and cloud at the same time to use the distributed system to the maximum of its potential.

\begin{figure}[t]
\centering
  \includegraphics[width=0.5\linewidth]{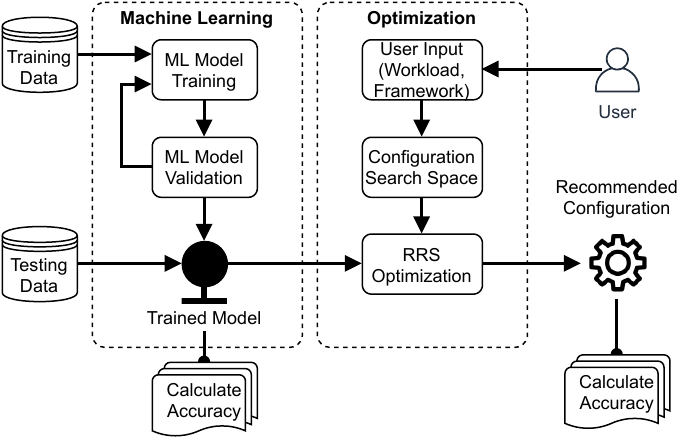}
\caption{The architecture of our co-tuning approach - \textit{TUNER}}
\label{tuner-architecture}
\end{figure}

However, manually co-tuning the cloud and data platform is a tedious and time-consuming task as described in Section \ref{background}. Therefore, we present \textit{TUNER} - an automated co-tuning approach that recommends cloud and platform configuration to reduce execution time and cost. The architecture of \textit{TUNER} is depicted in Figure \ref{tuner-architecture} and described in detail in the following sections. \textit{TUNER} consists of two phases - offline and online. First, in the offline phase, \textit{TUNER} uses Machine Learning (ML) to learn the performance model of Hadoop, Spark, and Flink for various workloads. Second, in the online phase, a user submits requirements (platform and workload) to \textit{TUNER} and then \textit{TUNER} uses the performance model and Recursive Random Search (RRS) algorithm to recommend cloud configuration as well as platform configuration for executing the workload.

\subsection{Performance Modelling} In order to co-tune cloud and data platform, we first need to learn how cloud and data platforms perform with respect to various workloads. We choose ML modelling for such learning. We leverage the machine learning approach to capture complex system dynamics that other techniques (e.g., rule-based and simulation-based) are unable to capture. Also, using machine learning enables us to utilize our learning obtained from real experiments reported in Section \ref{correlation}. ML models learn from previous data referred to as training data, which needs to be collected a priori. 

\subsubsection{Data Collection} Collecting high-quality training data for performance modeling in the context of cloud and data platforms is a challenging task \cite{herodotou2020survey}, \cite{costa2021survey}. This is because (i) the search space consists of a very large number of potential configurations for cloud and data platforms and (ii) executing each configuration to collect data is quite time-consuming. Therefore, collecting data for each configuration is almost impractical. In comparison to previous studies (e.g., \cite{yigitbasi2013towards, wang2016novel, wang2012predator, herodotou2011starfish, wang2012predator, herodotou2011no, chen2013cresp, fu2015drs}), we selected a larger set of potential cloud configurations and platform configurations as presented in Table \ref{hadoop_parameters_table}, Table \ref{spark_parameters_table}, Table \ref{flink_parameters_table}, and Table \ref{cloud_configurations_table}. Our training data consists of 1881 data points. These points come from the three platforms - Hadoop (11 $\times$ 20 $\times$ 3) + Spark (11 $\times$ 20 $\times$ 3) + Flink (11 $\times$ 17 $\times$ 3) = 1881. In (A $\times$ B $\times$ C),  \textit{A} denotes the number of cloud configurations, \textit{B} denotes the number of platform configurations, and \textit{C} denotes the number of workloads. For each of these data points, we executed the respective combination of cloud configuration, platform configuration, and workload. Then, we extracted the execution time from the logs and used it as a label for the respective data point. Therefore, each of our training data point consisted of cloud configuration, platform configuration, workload, and execution time (label). The generalization of our training model depends upon the diversity of the workloads. Whilst initially our ML model is trained for the three workloads, it can be repeated on regular intervals in future to incorporate new workloads. 

\subsubsection{ML Modelling} 
\begin{figure*}
\centering
\begin{subfigure}{.30\textwidth}
  \centering
  \includegraphics[width=\linewidth]{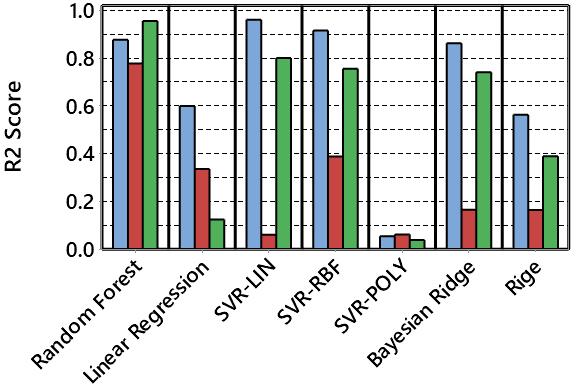}
  \caption{Hadoop}
  \label{fig:Flink-Resource-CPU}
\end{subfigure}%
\begin{subfigure}{.30\textwidth}
  \centering
  \includegraphics[width=\linewidth]{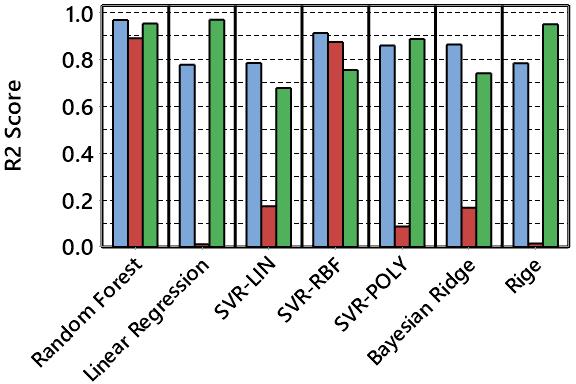}
  \caption{Spark}
  \label{fig:Flink-Resource-Memory}
\end{subfigure}
\begin{subfigure}{.30\textwidth}
  \centering
  \includegraphics[width=\linewidth]{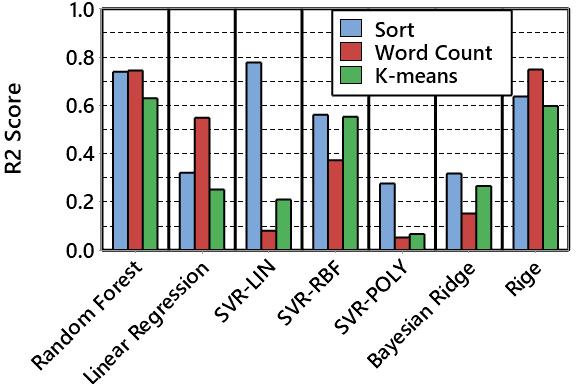}
  \caption{Flink}
  \label{fig:Flink-Resource-Disk}
\end{subfigure}
\caption{Prediction Accuracy of \textit{TUNER} Performance Model with various machine learning algorithms}
\label{rootmeansqureerror}
\end{figure*}

As shown in Figure \ref{tuner-architecture}, we used the training data to build an ML-based performance model. For ML model building, we have several ML algorithms available. To select the most accurate algorithm, we first tried seven algorithms i.e., random forest, linear regression, SVR-LIN, SVR-RBF, SVR-POLY, Bayesian ridge, and rige. For model training, we split the data into 70\% and 30\%, where 70\% data was used for training and 30\% was used for validation. We measured the validation accuracy using R2 score \cite{vidyullatha2016machine}, which measures how close the actual values are to the predicted values for data samples (i.e., the execution time for a given cloud + data platform + workload). In other words, the R2 score represents the prediction power of the model - a value closer to 1 indicates a more accurate prediction. The R2 scores we obtained for various ML algorithms are presented in Figure \ref{rootmeansqureerror}, which indicates that the scores for random forest are closer to 1. Therefore, we selected the performance model based on random forest for recommending configurations for cloud and data platforms.

\subsection{Configuration Tuning} We can use the ML model trained in the previous step to predict the execution time for a given set of cloud configuration, platform configuration, and workload. However, the model is trained on a specific set of sample data points. Therefore, the model is not representative of the entire search space. Taking inspiration from state-of-the-art \cite{herodotou2011profiling, herodotou2011no, herodotou2011starfish}, we used an optimization algorithm - Recursive Random Search (RRS) to reduce the search space. Due to its root in random sampling, RRS is robust to noise and quite suitable for parameter configuration problems \cite{herodotou2011profiling, herodotou2011no, herodotou2011starfish}. RRS algorithm first selects random samples from the global search space to identify promising regions. Then the algorithm leverages the trained ML model to recursively search for local optimum as depicted in Figure \ref{tuner-architecture}. RRS algorithm performs random sampling multiple times and each time the algorithm moves or shrinks based on previous sampling. Essentially, RRS follows a black-box approach, where it looks at high-level features in the initial phase of sampling and gradually gets deeper and deeper as sub-spacing happens before it finally converges to local optimums.  Once RRS recommends cloud and platform configuration for a given workload, \textit{TUNER} submits the job to the master node of the cluster along with the recommended configuration. Consequently, the master node executes the job with the tuned/recommended configurations. 

\subsection{Performance Evaluation} We now evaluate the performance of \textit{TUNER} in terms of its prediction accuracy and ability to reduce execution time and cost for a given data processing job. 

\subsubsection{TUNER's Recommended Configurations} Table \ref{hadoop-tuned_parameters_table}, Table \ref{spark-tuned_parameters_table}, and Table \ref{flink-tuned_parameters_table} show the cloud and platform configurations recommended by \textit{TUNER} for given workloads. The platform configurations recommended by \textit{TUNER} are quite different than the default configurations of Hadoop, Spark, and Flink presented in Table \ref{hadoop_parameters_table}, Table \ref{spark_parameters_table}, and Table \ref{flink_parameters_table}. Similarly, the recommended cloud configurations also vary across workloads. As an example, for executing Sort with Hadoop, \textit{TUNER} recommends cloud configuration \textit{C2}. On the other hand, \textit{TUNER} recommends cloud configuration \textit{C6} and \textit{C9} for Word Count and K-means, respectively. The recommended configurations even vary with respect to the platforms. For instance, to execute Sort, \textit{TUNER} recommends three different cloud configurations (\textit{C2}, \textit{C7}, and \textit{C4}) for the three platforms i.e., Hadoop, Spark, and Flink. The platform configurations also vary with respect to the workloads. Taking the example of \textit{spark.io.compression.codec}, \textit{TUNER} recommends to tune it to \textit{Snappy} for Word Count while recommends to tune it to \textit{Iz4} and \textit{Izf} for K-means and Sort, respectively. Our finding implies that the configuration suitable for one workload or platform is not necessarily suitable for running another type of workload or running the same workload on another platform. This also highlights the need for an automated tuning approach as manual tuning in a dynamic environment where workload and platform changes can become too tedious and time-consuming.

\begin{table}[]
\caption{\textbf{Hadoop} parameter configurations and cloud configuration recommended by \textit{TUNER} for each workload. \textit{C2}, \textit{C6}, and \textit{C9} specify cloud configurations presented in Table \ref{cloud_configurations_table}}
\centering
\scriptsize
%\resizebox{\textwidth}{!}{%
\begin{tabular}{l|l|ccc}
\hlineB{3}
\multirow{2}{*}{\textbf{ID}} & \multicolumn{1}{c|}{\multirow{2}{*}{\textbf{Hadoop Parameter}}} & \multicolumn{3}{c}{\textbf{Workload}} \\ \cline{3-5} 
 & \multicolumn{1}{c|}{} & \multicolumn{1}{c|}{Sort} & \multicolumn{1}{c|}{Word Count} & K-means \\ \hlineB{3}
H1 & output.fileoutputformat.compress & \multicolumn{1}{c|}{FALSE} & \multicolumn{1}{c|}{TRUE} & FALSE \\ 
H2 & output.fileoutputformat.compress.type & \multicolumn{1}{c|}{BLOCK} & \multicolumn{1}{c|}{RECORD} & BLOCK \\ 
H3 & output.fileoutputformat.compress.codec & \multicolumn{1}{c|}{GZip} & \multicolumn{1}{c|}{GZip} & BZip2 \\ 
H4 & output.compress & \multicolumn{1}{c|}{FALSE} & \multicolumn{1}{c|}{FALSE} & TRUE \\ 
H5 & map.output.compress.codec & \multicolumn{1}{c|}{GZip} & \multicolumn{1}{c|}{GZip} & GZip \\ 
H6 & tasktracker.map.tasks.maximum & \multicolumn{1}{c|}{4} & \multicolumn{1}{c|}{2} & 2 \\ 
H7 & tasktracker.reduce.tasks.maximum & \multicolumn{1}{c|}{4} & \multicolumn{1}{c|}{1} & 1 \\ 
H8 & child.java.opts & \multicolumn{1}{c|}{-Xmx200m} & \multicolumn{1}{c|}{-Xmx200m} & -Xmx200m \\
H9 & map.speculative & \multicolumn{1}{c|}{TRUE} & \multicolumn{1}{c|}{TRUE} & FALSE \\
H10 & reduce.speculative & \multicolumn{1}{c|}{TRUE} & \multicolumn{1}{c|}{TRUE} & FALSE \\ 
H11 & task.io.sort.mb & \multicolumn{1}{c|}{100} & \multicolumn{1}{c|}{10} & 100 \\ 
H12 & task.io.sort.factor & \multicolumn{1}{c|}{0.3} & \multicolumn{1}{c|}{0.5} & 0.8 \\
C & Cloud Configuration & \multicolumn{1}{c|}{C2} & \multicolumn{1}{c|}{C6} & C9 \\ \hlineB{3}
\end{tabular}%
%}
\label{hadoop-tuned_parameters_table}
\end{table}

\begin{table}[]
\caption{\textbf{Spark} parameter configurations and cloud configuration recommended by \textit{TUNER} for each workload. \textit{C7}, \textit{C8}, and \textit{C4} specify cloud configurations presented in Table \ref{cloud_configurations_table}}
\centering
\scriptsize
%\resizebox{\textwidth}{!}{%
\begin{tabular}{l|l|ccc}
\hlineB{3}
\multirow{2}{*}{\textbf{ID}} & \multicolumn{1}{c|}{\multirow{2}{*}{\textbf{Spark Parameter}}} & \multicolumn{3}{c}{\textbf{Workload}} \\ \cline{3-5} 
 & \multicolumn{1}{c|}{} & \multicolumn{1}{c|}{Sort} & \multicolumn{1}{c|}{Word Count} & K-means \\ \hlineB{3}
S1 & io.compression.coded & \multicolumn{1}{c|}{Izf} & \multicolumn{1}{c|}{Snappy} & Iz4 \\
S2 & serializer & \multicolumn{1}{c|}{Kryo} & \multicolumn{1}{c|}{Kryo} & Java \\ 
S3 & io.compression.Iz4.blockSize & \multicolumn{1}{c|}{64k} & \multicolumn{1}{c|}{64k} & 64k \\ 
S4 & shuffle.spill.compress & \multicolumn{1}{c|}{TRUE} & \multicolumn{1}{c|}{FALSE} & TRUE \\ 
S5 & reducer.maxSizeInFlight & \multicolumn{1}{c|}{48m} & \multicolumn{1}{c|}{72m} & 72m \\ 
S6 & shuffle.file.buffer & \multicolumn{1}{c|}{16k} & \multicolumn{1}{c|}{48k} & 16k \\ 
S7 & shuffle.compress & \multicolumn{1}{c|}{TRUE} & \multicolumn{1}{c|}{FALSE} & TRUE \\ 
S8 & broadcast.blockSize & \multicolumn{1}{c|}{4m} & \multicolumn{1}{c|}{2m} & 2m \\ 
S9 & locality.wait & \multicolumn{1}{c|}{1s} & \multicolumn{1}{c|}{3s} & 1s \\ 
S10 & memory.fraction & \multicolumn{1}{c|}{0.4} & \multicolumn{1}{c|}{0.6} & 0.4 \\ 
S11 & memory.storageFraction & \multicolumn{1}{c|}{0.3} & \multicolumn{1}{c|}{0.7} & 0.7 \\ 
C & Cloud configuration & \multicolumn{1}{c|}{C7} & \multicolumn{1}{c|}{C8} & C4 \\ \hlineB{3}
\end{tabular}%
%}
\label{spark-tuned_parameters_table}
\end{table}

\begin{table}[!]
\caption{\textbf{Flink} parameter configurations and cloud configuration recommended by \textit{TUNER} for each workload. \textit{C7}, \textit{C10}, and \textit{C4} specify cloud configurations presented in Table \ref{cloud_configurations_table}}
\centering
\scriptsize
%\resizebox{\textwidth}{!}{%
\begin{tabular}{l|l|ccc}
\hlineB{3}
\multirow{2}{*}{\textbf{ID}} & \multicolumn{1}{c|}{\multirow{2}{*}{\textbf{Flink Parameter}}} & \multicolumn{3}{c}{\textbf{Workload}} \\ \cline{3-5} 
 & \multicolumn{1}{c|}{} & \multicolumn{1}{c|}{Sort} & \multicolumn{1}{c|}{Word Count} & K-means \\ \hlineB{3}
F1 & memory.managed.fraction & \multicolumn{1}{c|}{0.5} & \multicolumn{1}{c|}{0.4} & 0.3 \\ 
F2 & memory.jvm-overhead.fraction & \multicolumn{1}{c|}{0.05} & \multicolumn{1}{c|}{0.05} & 0.2 \\ 
F3 & memory.network.fraction & \multicolumn{1}{c|}{0.5} & \multicolumn{1}{c|}{0.5} & 0.5 \\ 
F4 & network.blocking-shuffle.compression.enabled & \multicolumn{1}{c|}{TRUE} & \multicolumn{1}{c|}{TRUE} & FALSE \\ 
F5 & network.memory.buffers-per-channel & \multicolumn{1}{c|}{2} & \multicolumn{1}{c|}{2} & 4 \\ 
F6 & network.netty.server.numThreads & \multicolumn{1}{c|}{2} & \multicolumn{1}{c|}{1} & 2 \\ 
F7 & network.netty.clinet.num-threads & \multicolumn{1}{c|}{2} & \multicolumn{1}{c|}{1} & 2 \\ 
F8 & execution.checkpointing.snapshot-compression & \multicolumn{1}{c|}{TRUE} & \multicolumn{1}{c|}{TRUE} & TRUE \\ 
C & Cloud Configuration & \multicolumn{1}{c|}{C4} & \multicolumn{1}{c|}{C10} & C4 \\ \hlineB{3}
\end{tabular}%
%}
\label{flink-tuned_parameters_table}
\end{table}

\subsubsection{TUNER Prediction Accuracy and Performance Improvement} 
\begin{figure*}[!tbp]
\centering
\begin{subfigure}{.33\textwidth}
  \centering
  \includegraphics[trim=0 0 0 0, clip, scale = 0.35]{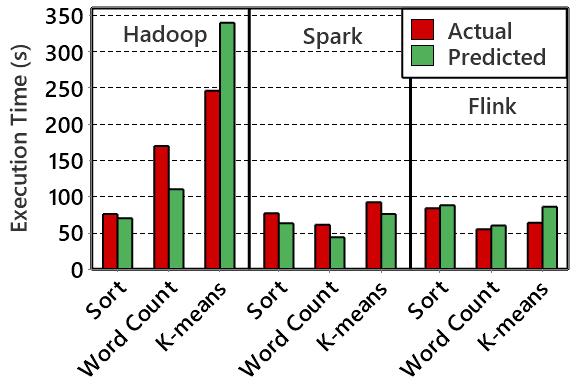}
  \caption{Execution time - Actual VS Predicted}
  \label{actual_predicted_time}
\end{subfigure}%
\begin{subfigure}{.33\textwidth}
  \centering
  \includegraphics[trim=0 0 0 0, clip, scale = 0.35]{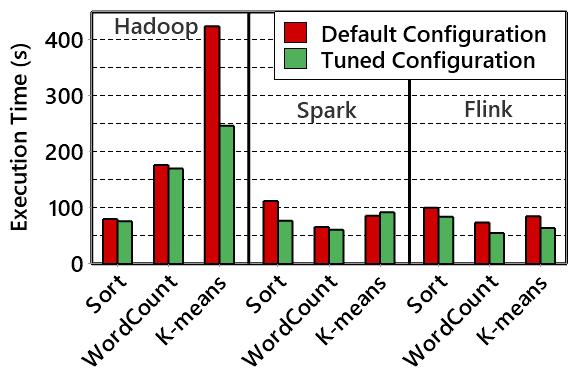}
  \caption{Execution time - Default VS Tuned}
  \label{default_tuned_time}
\end{subfigure}
\begin{subfigure}{.3\textwidth}
  \centering
  \includegraphics[trim=0 0 0 0, clip, scale = 0.35]{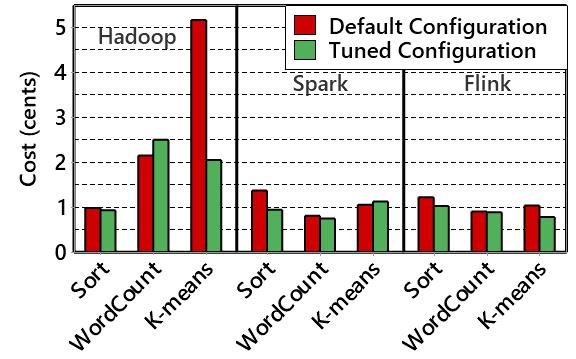}
  \caption{Cost - Default VS Tuned}
  \label{default_tuned_cost}
\end{subfigure}
\caption{\textit{TUNER's} impact on execution time and cost.}
\label{improvement_figure}
\end{figure*}

After \textit{TUNER} recommended the cloud and platform configuration for various workloads, we executed each workload with the recommended configuration. Figure \ref{actual_predicted_time} shows the actual and predicted execution times for the configurations (Table \ref{hadoop-tuned_parameters_table}, Table \ref{spark-tuned_parameters_table}, and Table \ref{flink-tuned_parameters_table}) recommended by \textit{TUNER}. Whilst predicted execution time is the time predicted by \textit{TUNER} for the recommended configuration, actual execution times are calculated based on the real execution of the workload with the recommended cloud and platform configuration. We observe that \textit{TUNER} can accurately predict the execution time of a workload for a given cloud and platform configuration. The mean relative error between actual and predicted execution time is 15.6\%.  Figure \ref{default_tuned_time} and Figure \ref{default_tuned_cost} respectively show the execution time and cost with default and tuned configurations. The cost is calculated based on the prices shown in Table \ref{amazon_azure_google} for various VMs. As compared to the default configuration, the execution time and cost are reduced for almost all workloads and platforms. We considered data platforms' default configurations as the performance baseline and calculated the performance gain of \textit{TUNER} for each cloud configuration (i.e., \textit{C0-C10}). Overall, \textit{TUNER} reduces the mean execution time by 17.5\% and mean cost by 14.9\% compared to the baseline performances of these platforms, the mean reduction in execution time is 19.7\%, 10.6\%, and 22.2\% for Hadoop, Spark, and Flink, respectively. This trend is aligned with the variance in execution time with respect to various cloud and platform configurations presented in Figure \ref{hadoop_execution_time}, Figure \ref{spark_execution_time}, and Figure \ref{flink_execution_time}. Wherever the variance is high, \textit{TUNER} has more opportunity to exploit the configurations for improving the execution time. For example, the variance in execution time of Hadoop with various cloud and platform configurations is high as compared to Spark. Therefore, \textit{TUNER} has a better chance to improve the execution time of Hadoop as compared to the little improvement available for Spark. Another potential reason for the lesser reduction with Spark is the excessive garbage collection by Spark in memory \cite{marcu2016spark}, which does not let Spark fully exploit the underlying platform configurations.

\section{Conclusion}
Distributed data processing systems leverage data platforms (e.g., Hadoop, Spark, and Flink) to distribute the storage and processing of data among computing nodes of a cloud. These data platforms come with several configuration parameters that require tuning in order to use the distributed system to the maximum of its potential. Furthermore, these distributed systems are deployed on a cloud, which also requires tuning to determine optimal configuration for various aspects such as the number and type of virtual machines on the cloud. In this paper, we first studied the impact of cloud configuration on data platform configuration. After understanding the impact, we proposed a co-tuning technique, \textit{TUNER}, that leverages machine learning and RRS optimization to recommend configurations for cloud and data platforms for efficient usage of distributed systems in a cloud environment. For experimentation, we used OpenStack cloud and the three most widely used data platforms - Hadoop, Spark, and Flink. Moreover, we conducted the experiments with three benchmarking workloads i.e., Sort, Word Count, and K-means. Our findings reveal the following.
\begin{itemize}

\item We found that cloud configuration impacts data platform configuration. In other words, the way we configure cloud significantly impacts the way we configure a data platform. 
\item Cloud configurations more significantly impact the execution time of a distributed system as compared to the impact of the configuration of data platform on execution time.
\item We observed that Hadoop, Spark, and Flink perform better in homogeneous cluster (where all nodes are of the same type) as compared to heterogeneous cluster (where nodes are of different types). Also, Hadoop performs better in a cloud cluster having more nodes of small size while Flink performs better in a cluster having few nodes of large size. 
\item With respect to default settings,  our co-tuning approach (\textit{TUNER}) reduces the execution time and cost by 17.5\% and 14.9\%, respectively. Among the seven studied machine learning algorithms for \textit{TUNER}, we found random forest as the most accurate in terms of performance prediction.

\end{itemize}

The work reported in this paper has implications both for practitioners and researchers. For practitioners, the key takeaway from this paper is not to treat cloud configuration and data platform configuration as silos rather co-tune the two so as to gain maximum performance improvement. Moreover, our findings suggest that the proposed co-tuning approach, \textit{TUNER}, can facilitate practitioners to automatically co-tune cloud and data platform. For researchers, our work has uncovered several areas for future research. Given that in addition to data platforms, data storage solutions such as Cassandra and MongoDB are also critical components of a distributed system. Thus, it will be fruitful to undertake a similar study for data storage solutions. Computing nodes of a cloud can have different storage types such as Hard Disk Drive (HDD) and Solid State Drive (SSD). It will be worth exploring how storage type of the nodes affect the configuration of cloud and the configuration of the data platform deployed on the cloud. Whilst our focus in this paper was on tuning data platforms for batch processing, an interesting avenue for future research is to correlate the tuning of cloud and data platforms for stream data processing, where data is processed on the fly in real-time. Finally, the placement of VMs on same or different physical machines is also a crucial factor in the context of cloud computing. Therefore, it should be investigated as to how the placement of VMs impact the configuration tuning of cloud as well as configuration of the deployed data platform.

\section*{Acknowledgments}
We acknowledge the support provided by Sharon Khate Damaso and Atukoralage Kusala Rajakaruna with regards to the execution of experiments. We also thank the anonymous reviewers for providing useful comments to help improve the manuscript.

\bibliography{bigdata}

\end{document}